\documentclass[pra,aps,twocolumn,10pt]{revtex4-1}
\usepackage{graphicx}
\usepackage{rotating}
\usepackage{amsmath}

\begin{document}

\title{Universal predictability of mobility patterns in cities}

\author{Xiao-Yong Yan$^{1,2}$}
\author{Chen Zhao$^{1}$}
\author{Ying Fan$^{1}$}
\author{Zengru Di$^{1}$}
\author{Wen-Xu Wang$^{1,3}$  }
\email{wenxuwang@bnu.edu.cn}
\affiliation{
$^{1}$School of Systems Science, Beijing Normal University, Beijing 100875, P.R. China\\
$^{2}$Department of Transportation Engineering, Shijiazhuang Tiedao University, Shijiazhuang 050043, P.R. China\\
$^{3}$School of Electrical, Computer and Energy Engineering, Arizona State University, Tempe, AZ 85287, United States
}

\begin{abstract}
Despite the long history of modelling human mobility, we continue to lack a highly accurate approach with low data requirements for predicting mobility patterns in cities. Here, we present a population-weighted opportunities model without any adjustable parameters to capture the underlying driving force accounting for human mobility patterns at the city scale. We use various mobility data collected from a number of cities with different characteristics to demonstrate the predictive power of our model. We find that insofar as the spatial distribution of population is available, our model offers universal prediction of mobility patterns in good agreement with real observations, including distance distribution, destination travel constraints and flux. In contrast, the models that succeed in modelling mobility patterns in countries are not applicable in cities, which suggests that there is a diversity of human mobility at different spatial scales. Our model has potential applications in many fields relevant to mobility behaviour in cities, without relying on previous mobility measurements.
\end{abstract}

\maketitle

\section{Introduction}
Predicting human mobility patterns is not only a fundamental problem in geography and spatial economics~\cite{bart}; it also has many practical applications in urban planning~\cite{batty}, traffic engineering~\cite{mt,helb}, infectious disease epidemiology ~\cite{huf,eub,bal}, emergency management~\cite{bagrow1,holme,rahwana} and location-based service~\cite{lbs}. Since the 1940s, many trip distribution models~\cite{io,gr,ru,radi,cont,rank,leno} have been presented to address this challenging problem, among which the gravity model is the prevailing framework~\cite{gr}. Despite its wide use  in predicting mobility patterns at different spatial scales~\cite{jung,kri,kal,goh}, the gravity model relies on specific parameters fitted from systematic collections of traffic data. If previous mobility measurements are lacking, the gravity model is not applicable. A similar limitation exists in all trip distribution models that rely on context-specific parameters, such as the intervening opportunity model~\cite{io}, the random utility model~\cite{ru} and others.

Quite recently, the introduction of the radiation model~\cite{radi} has provided a new insight into the long history of modelling population movement. The model is based on a solid theoretical foundation and can precisely reproduce observed mobility patterns ranging from long-term migrations to inter-county commutes. Surprisingly, the model needs only the spatial distribution of population as an input, without any adjustable parameters. Nevertheless, some evidence has demonstrated that the radiation model may be not applicable to predicting human mobility at the city scale~\cite{gvr,lx2}. Understanding mobility patterns in cities is of paramount importance in the sense that cities are the foci of disease propagation, traffic congestion and pollution~\cite{eub,city}, partly resulting from human movement. These problems can be resolved through developing more efficient transportation systems and optimising traffic management strategies, all of which depend on our ability to predict human travel patterns in cities~\cite{song}. Despite the success of the radiation model in countries, we continue to lack an explicit and comprehensive understanding of the underlying mechanism accounting for the observed mobility patterns in cities. We argue that this is mainly ascribed to the relatively high mobility of residents in cities compared to larger scales, such as travelling among counties. Inside cities, especially metropolises, high development of traffic systems allow residents to travel relatively long distance to locations with more opportunities and attraction. In this sense, the models that are quite successful in reproducing mobile patterns at large spatial scales fail at the city scale. Yet, revealing the underlying driving force and restrictions for such mobility to predict mobile patterns in cities remains an outstanding problem. 

\begin{figure*}
\begin{center}
\centerline{\includegraphics[width=180mm]{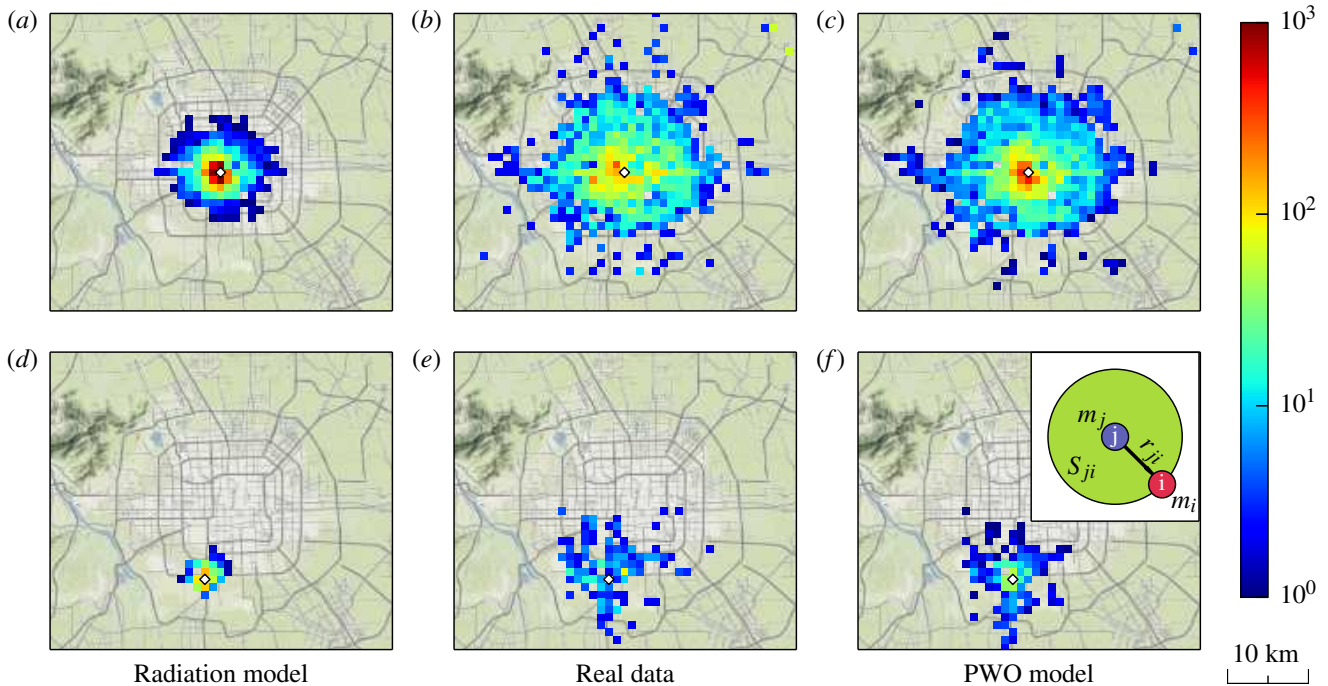}}
\caption{\label{fig-diff} Human mobility range at city scale. ({\it a-c}) Spatial distributions of destination selections for travelling  from a downtown location (displayed as a diamond-shaped dot) in Beijing. ({\it d-f}) Spatial  distributions of travel from a suburban location. From left to right, the panels correspond to the results generated by the radiation mode, the observed  data and the results generated by the PWO model. The colour  bar represents the number of travellers from the origin to a destination. The inset offers definitions of variables used in the model, in which the purple circle (location $i$) denotes an origin with population $m_i$, the blue circle (location $j$) stands for a destination with population $m_j$ and $S_{ji}$ represents the total population in a circular area of radius $r_{ji}$ centred at location $j$ (including $m_i$ and $m_j$).}
\end{center}
\end{figure*}

In this paper, we develop a population-weighted opportunities model without any adjustable parameters as an alternative to the radiation model to predict human mobility patterns in a variety of cities.  Insofar as the distribution of population in different cities are available, our model offers universal prediction of human mobility patterns in several cities as quantified by some key measurements, including distance distribution, destination travel constraints and flux. In contrast, the models that succeed in predicting mobility patterns at large spatial scales, such as countries, are inappropriate at the city scale because of the underestimation of human mobility. Our approach suggests the diversity of human mobility at different spatial scales, deepening our understanding of human mobility behaviours. 

~
\\
~
\\

\section{Results}
\subsection{Population-weighted opportunities model}
The model is derived from a stochastic decision making process of individual's destination selection. Before an individual selects a destination, s/he will weigh the benefit of each location's opportunities. The more opportunities a location has, the higher the benefit it offers and the higher the chance of it being chosen. Although the number of a location's opportunities is difficult to straightforwardly measure, it can be reflected by its population. Insofar as the population distribution is available, it is reasonable to assume that the number of opportunities at a location is proportional to its population, analogous to the assumption of the radiation model~\cite{radi}.

In contrast to the radiation model's assumption that individuals tend to select the nearest locations with relatively larger benefits, we enlarge the possible chosen area of individuals to include the whole city regarding the relatively high mobility at the city scale. As shown in figure~\ref{fig-diff}, our assumption leads to much better prediction than that of the radiation model. Nevertheless, the possibility of travel in the observed data still decays as the distance between origin and destination increases. Such decay, as predicted by different models and common in real observations, results from the reduction of attraction associated with a type of cost. For example, the gravity model~\cite{gr} assumes that the attraction of a destination, i.e., its opportunities, is reduced according to a function of the distance from the origin. However, the distance function inevitably includes at least one parameter. To capture the mobility behaviours and avoid adjustable parameters, we simply assume that the attraction of a destination is inversely proportional to the population $S_{ji}$ in the circle centred at the destination with radius $r_{ij}$ (the distance between the origin $i$ and destination $j$,  as illustrated in the inset of figure~\ref{fig-diff}), minus a finite-size correction $1/M$, \textit{i.e.}, 

\begin{figure}
\centerline{\includegraphics[width=90mm]{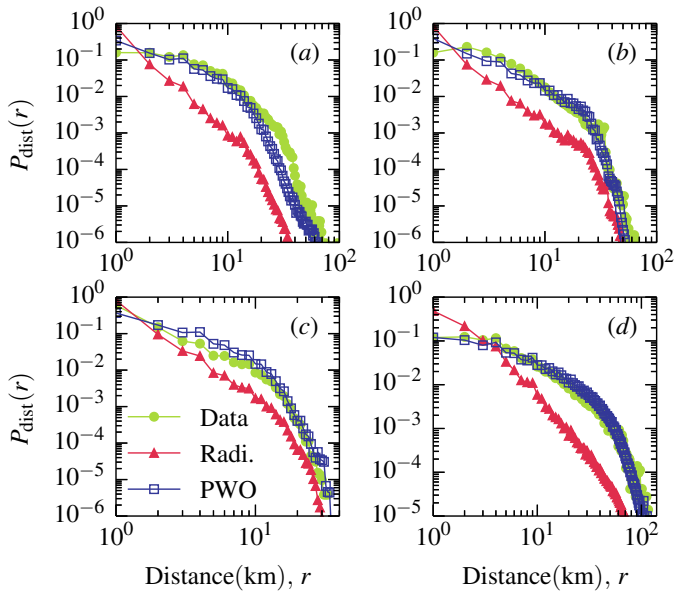}}
\caption{\label{fig-tdd} Travel distance distributions $P_{\rm{dist}}(r)$ produced by the radiation model and the PWO model in comparison with empirical data. Four cities, ({\it a}) Beijing, ({\it b}) Shenzhen, ({\it c}) Abidjan and ({\it d}) Chicago, are studied. Here, $P_{\rm{dist}}(r)$ is defined as the probability of  travel between locations at distance $r$.}

\end{figure}
\begin{figure}
\centerline{\includegraphics[width=90mm]{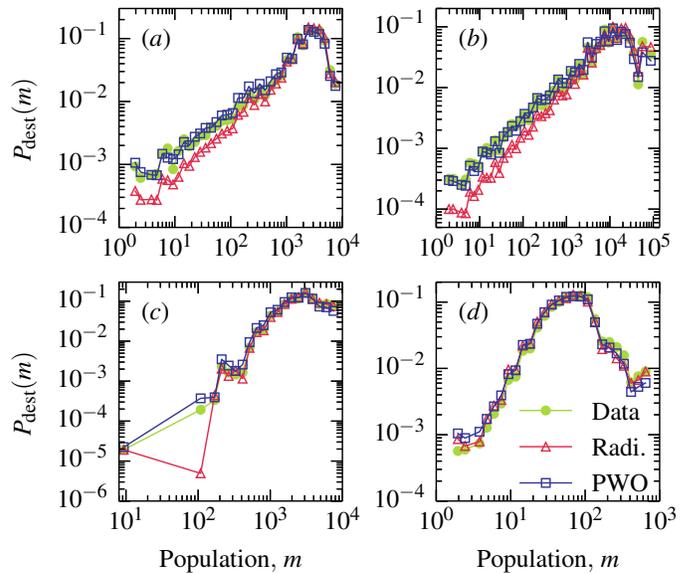}}
\caption{\label{fig-des} Comparing the destination travel constraints of the radiation and the PWO models with real data. Four cities  are explored as representative cases: ({\it a}) Beijing, ({\it b}) Shenzhen, ({\it c}) Abidjan and ({\it d}) Chicago.  $P_{\rm{dest}}(m)$ is the probability of travel towards a location with population $m$.}
\end{figure}
\begin{figure*}
\begin{center}
\centerline{\includegraphics[width=168mm]{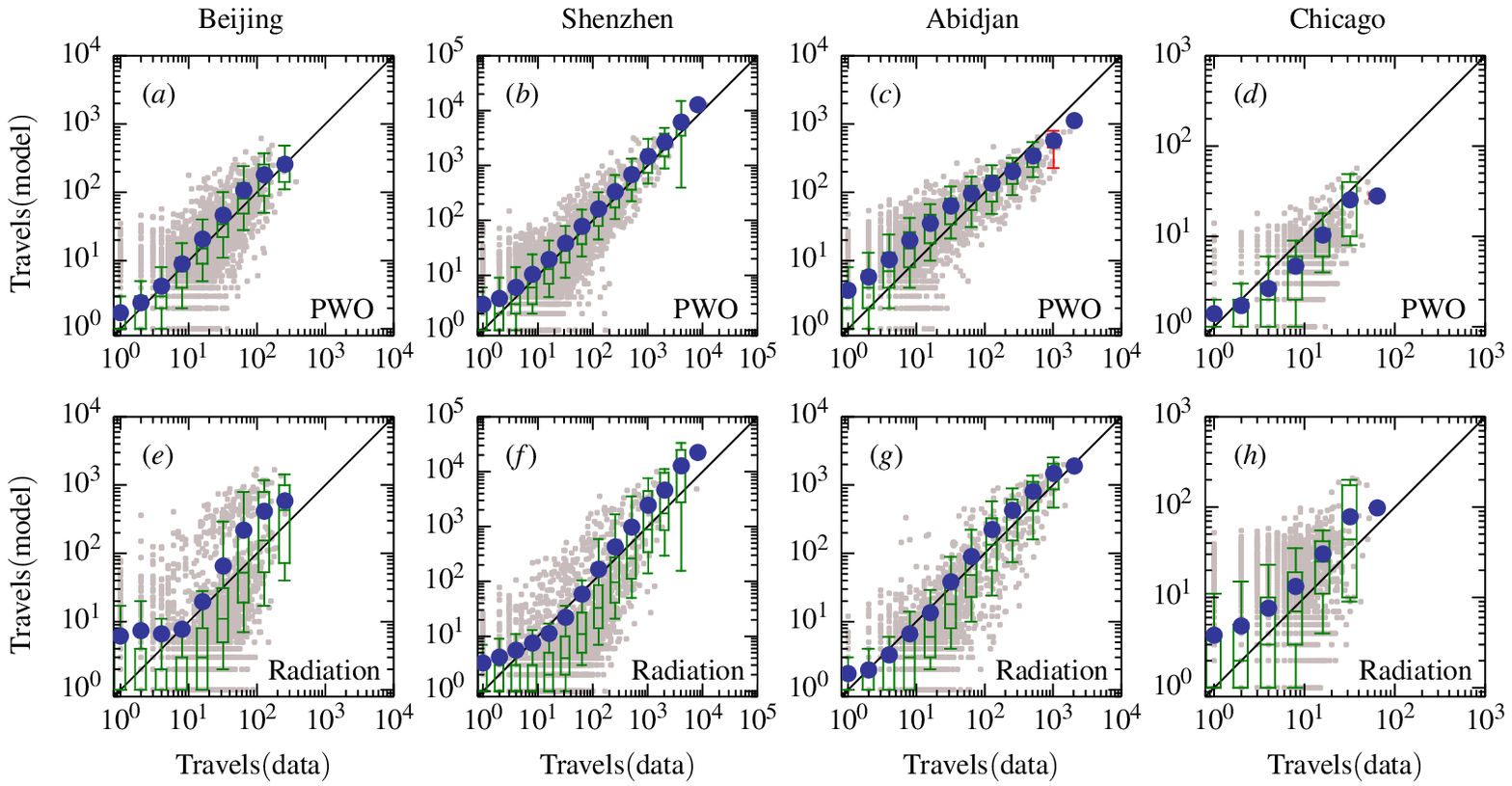}}
\caption{\label{fig-box} Comparing the observed fluxes with the predicted fluxes in four cities. ({\it a-d}) Travel fluxes predicted by the PWO model. ({\it e-h}) Travel fluxes predicted by the radiation model. The grey points are scatter plot for each pair of locations. The blue points represent the average number of predicted travels in different bins. The boxplots, obtained via standard statistical methods~\cite{boxplot}, represent the distribution of the number of predicted travels in different bins of the number of observed travels. A box is marked in green if the line $y=x$ lies between $10\%$ and $91\%$ in that bin and in red otherwise.
}
\end{center}
\end{figure*}
\begin{equation}
\label{eqat}
A_j=o_j(\frac{1}{S_{ji}}-\frac{1}{M})
\end{equation}
where $A_j$ is the relative attraction of destination $j$ to travellers at origin $i$, $o_j$ is the total opportunities of destination $j$ and $M$ is the total population in the city. Further, assuming that the probability of travel from $i$ to $j$ is proportional to the attraction of $j$ and recalling the assumption that the number of opportunities $o_j$ is proportional to the population $m_j$, we have the travel from  $i$ to $j$ as 
\begin{equation}
\label{eqcd}
T_{ij}=T_{i}\frac{m_{j}(\frac{1}{S_{ji}}-\frac{1}{M})}{\sum^{N}_{k \neq i}{m_{k}(\frac{1}{S_{ki}}-\frac{1}{M})}},
\end{equation}
where $T_{i}$ is the trips departing from $i$ and $N$ is the number of locations in the city.

The presented model reflects the effect of competition for opportunities among  potential destinations: for a traveller at origin $i$ travelling to a potential destination $j$, more population between $i$ and $j$ will induce stronger competition for limited opportunities, so that the probability of being offered opportunities will be lower. In this regard, it is reasonable to assume that the attraction of a destination for a traveller is the destination's opportunities inversely weighted by population between the destination and the origin. We therefore name our model the \textit{population-weighted opportunities model} (PWO). We will then demonstrate the universal predictability of mobility patterns in cities via the PWO model through  a variety of real travel data in several cities.

\subsection{Predicting mobility patterns}
To validate the PWO model by comparison with the performance of the radiation model(see details in {\bf Materials and methods, The radiation model}), we employ human daily travel data from four cities collected by GPS, mobile phone and traditional household surveys (see details in  {\bf Materials and methods, Data sets and Data preprocessing}).

Figure~\ref{fig-diff} exemplifies travel from a downtown and a suburban location in Beijing predicted in an intuitive manner by the PWO model and the radiation model in comparison with real data. It shows  that the radiation model underestimates the travel areas in both cases, whereas the travel patterns resulting from our model are quite consistent with empirical evidence, demonstrating the relatively higher mobility in cities than  at larger spatial scales where the radiation model succeeds, such as countries. 

We systematically investigate the travel distance distribution obtained by both models based on real data. Travel distance distribution is an important statistical property to capture human mobility behaviours~\cite{broc,gonz,rth} and reflect a city's economic efficiency ~\cite{bart}. We find that, as shown in figure~\ref{fig-tdd}, the  distributions of travel distance predicted  by the PWO model are  in good agreement with the real distributions. In contrast, the radiation model underestimates long-distance (longer than approximately 2 km) travel in all cases. This implies that the assumption of the radiation model is inappropriate at the city scale by precluding individuals from choosing relatively long travels to find better locations with more opportunities. 
The success of the PWO model in predicting real travel distance distributions in cities provides strong evidence for the validity of its basic assumptions.

We next explore the probability of travel towards a location with population $m$, say, $P_{\rm{dest}}(m)$, for both observed data and the predictive models. $P_{\rm{dest}}(m)$ is a key quantity  for measuring  the accuracy of origin-constrained mobility models (the radiation model and PWO model used here are both origin-constrained), because origin-constrained models cannot ensure the agreement between modelled travel to a location and real travel to the same location~\cite{mt}. In figure~\ref{fig-des}, we can see that our model equally or better predicts empirical observations compared to the radiation model.

A more detailed measure of a model's ability to predict mobility patterns can be implemented in terms of the travel fluxes between all pairs of locations produced by a model in comparison with real observations, as has been used in Ref.~\cite{radi}. As shown in figure~\ref{fig-box}, we find that---excepting the case of Abidjan---the average fluxes predicted by the radiation model deviate from the real fluxes, whereas the results from the PWO model are in reasonable agreement with real observations.

Note that the boxplot method used here cannot allow an explicit comparison to distinguish the performance of the two models. For example, in figure~\ref{fig-box}(e), although the results deviate from the empirical data significantly, the boxes are still coloured  green, suggesting the need of an alternative statistical method. Thus we exploit the S{\o}rensen similarity index~\cite{sorn}  (see details in {\bf Materials and methods, S{\o}rensen similarity index}) to quantify the degree of similarity with real observations to offer a better comparison. We have also applied both models to six European cities and another four U. S. cities to make a more comprehensive comparison (details are available in the {\bf Supplementary Material, Section S1} and {\bf S2}). The results are shown in figure~\ref{fig-ssi}. For all studied cases, our model outperforms the radiation model and exhibits relatively high index values, say, approximately 0.7, indicating that the PWO model captures the underlying mechanism that drives human movement in cities.

\section{Discussion}

We developed a population-weighted opportunities model as an alternative to the radiation model to reproduce and predict mobile behaviours in cities with different sizes, economic levels and cultural backgrounds. Our model needs only the spatial distribution of population as an input, without any adjustable parameters. The mobility patterns resulting from the model are in good agreement with real data with respect to travel distance distribution, destination travel constraints and flux, suggesting that the model captures the fundamental mechanisms governing human daily travel behaviours at city scale. 

The radiation model, despite having the advantage of being parameter-free and performing well at large spatial scale, cannot offer satisfactory predictions of mobility patterns at the city scale. The problem lies in the underestimation of the relatively high mobility at the city scale. In particular, the radiation model assumes that limited mobility prevents people from selecting a farther location with more opportunities to gain more benefits than a nearby location. This assumption is reasonable at the inter-city scale, but inappropriate in cities. The PWO model can successfully overcome this problem by assuming the attraction of a potential destination is inversely proportional to its population, which results from competition for opportunities in the whole city. Insofar as only population distribution is available, our model presently offers the best prediction of mobility patterns at the city scale, significantly deepening our understanding of human mobility in cities and demonstrating the universal predictability of mobility patterns at the city scale. 

\begin{figure}
\centerline{\includegraphics[width=90mm]{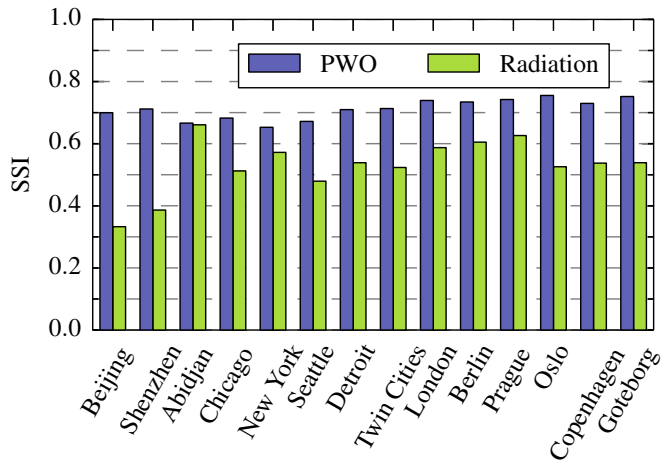}}
\caption{\label{fig-ssi} Comparison between the prediction ability of the PWO and the radiation model in terms of S{\o}rensen similarity index (SSI). SSI is calculated from Eq.~(\ref{eqssi}). 14 cities around the world are studied. }
\end{figure}

We have also compared the PWO model with three classical parameterised models: the gravity model~\cite{gr}, the intervening opportunity model~\cite{io} and the rank-based model~\cite{rank} (see details in the {\bf Supplementary Material, Section S3}). Although in rare cases the parameterised models can yield better predictive accuracy than the PWO model, their parameter-dependence nature limits their scope to the cases with particular previous systematic mobility measurements and relatively stable mobility patterns. The PWO model, without such limitations, has much more predictive power.  By exploring the relationship between all the previous models and our PWO model (see details in the {\bf Supplementary Material, Section S4}), we find that although these models have different hypotheses, they share an underlying mechanism: the probability that an individual selects a location to travel is decreased along with the increment of some prohibitive factors (distance or population). The key difference lies in that the gravity model, the intervening opportunity model and the rank-based model need adjustable parameters to quantify the decrement, whereas the decrement is naturally determined by population distribution in the radiation model and the PWO model. 

It is noteworthy that despite the advantages of the PWO model in predicting mobility patterns at the city scale, the predictability could
be improved further. The travel matrices established by the model share approximately  70\% common part with the real data (see figure~\ref{fig-ssi}). Although such accuracy can suffice for the requirements in many areas of applications, {\it e.g.}, in urban planning and epidemic modelling~\cite{eva}, it is still below the average upper limit of the predictability of human mobility~\cite{song}. In principle, the PWO model is essentially a type of aggregate travel model~\cite{mt} based on the collective behaviours of groups of similar travellers, whereas the diversity of real individuals' behaviours~\cite{yan,bagrow} is in contrast to the assumptions of aggregate models, accounting for their inaccuracy in reproducing and predicting movement patterns.  Microscopic mobility models, such as agent-based models~\cite{songnp,han,hasan,szell,schneider}may offer better prediction of mobility patterns as an alternative but suffer from much higher computational complexity. Therefore, an efficient macroscopic mobility model taking the diversity of individual behaviours into account would be worth pursuing in the future to further deepen our understanding of human mobility.

\section{Materials and methods}

\subsection{Data sets}

(1) Beijing taxi passengers. This data set is the travel records of taxi passengers in Beijing in a week~\cite{lx1}. When a passenger gets on or gets off a taxi, the coordinates and time are recorded automatically by a GPS-based device installed in the taxi. From the data set we extract 1,070,198 taxi passengers travel records. Some evidence indicates that in Beijing, the average travel distance of taxi passengers is similar to the commuting distance~\cite{qin14} and the spatial distribution of taxi passengers is similar to that of populations~\cite{yang14}. Thus the taxi passengers data can capture the travel pattern of urban residents to some extent, although taxi passengers only constitute a small subset of the population in a city.

\begin{figure}
\centerline{\includegraphics[width=90mm]{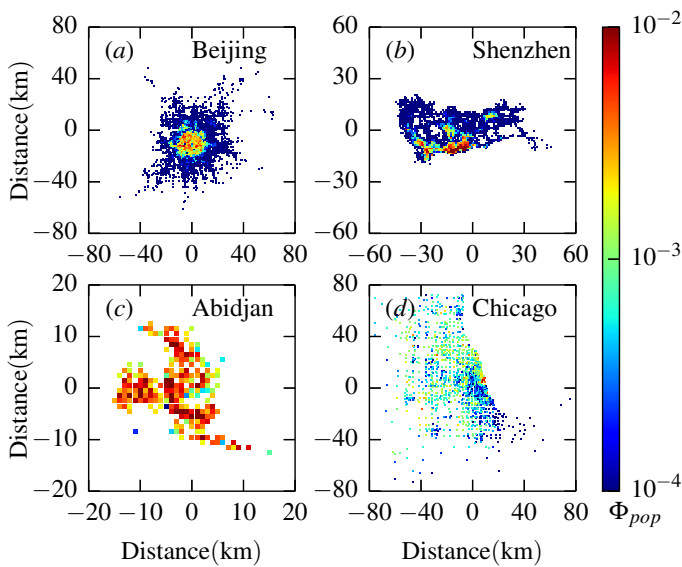}}
\caption{\label{fig-map} Zone partition and population density distribution in four cities. ({\it a}) Beijing with 3,025 zones, ({\it b}) Shenzhen with 1,634 zones, ({\it c}) Abidjan with 219 zones and ({\it d}) Chicago with 1,406 zones. The density function $\Phi _{pop}(i)$ represents the probability of finding travel starting from zone $i$. }
\end{figure}
(2) Shenzhen taxi passengers. The Shenzhen taxi passenger tracker data has the same data format as that of Beijing. The data set records 2,338,576 trips by taxi passengers in 13,798 taxis in Shenzhen from 18 Apr. 2011 to 26 Apr. 2011. 

(3) Abidjan mobile phone users. The data set contains 607,167 mobile phone users' movements between 381 cell phone antennas in Abidjan, the biggest city of Ivory Coast, during a two-week observation period~\cite{mobdata}. Each movement record contains the coordinates (longitude and latitude) of the origin and destination. The data set is based on the anonymised Call Detail Records (CDR) of phone calls and SMS exchanges between five million of Orange Company’s  customers in Ivory Coast. To protect customers' privacy, the customer identifications have been anonymised by Orange Company.

(4) Chicago travel tracker survey. Chicago travel tracker survey was conducted by the Chicago Metropolitan Agency for Planning during 2007 and 2008, which provides a detailed travel inventory for each member of 10,552 households in the greater Chicago area. The survey data are available online at  http://www.cmap.illinois.gov/travel-tracker-survey. Because  some participants provided one-day travel records but others provided two-days, to maintain consistency, we only extracted the first-day travel records from the data set. The extracted data include 87,041 trips, each of which includes coordinates of the trip's origin and destination.

\subsection{Data preprocessing} The raw travel data of four cities contain latitude and longitude coordinates of each traveller’s origin and destination. The raw data cannot be immediately used in mobility models. Alternatively, we used coarse-grained travel data through partitioning a city into a number of zones, each of which corresponds to a location in the literature~\cite{mt}. Because of the absence of natural partitions in cities (in contrast to states or counties), we simply partition all cities into equal-area square zones, each of which is of dimension 1 km $\times$ 1 km. Figure~\ref{fig-map} shows the zone partition results and the number of zones in four cities. We assign an origin (or destination) zone ID to each trip if its origin (or destination) falls into the range of that zone. Then, we can accumulate the total number $T_i$ of trips departed from an arbitrary zone $i$, and the total number $T_{ij}$ of trips  from zone $i$ to zone $j$. In general, the number of trips  departed from a zone is proportional to the population of the zone~\cite{radi}. The spatial distributions of population density estimated from travel data in the four cities are shown in figure~\ref{fig-map}.

\subsection{The radiation model}
The radiation model~\cite{radi} is a parameter-free model to predict travel fluxes among different locations based on population distribution:
\begin{equation}
\label{eqra}
T_{ij}=T_{i}\frac{m_{i}m_{j}}{(m_{i}+s_{ij})(m_{i}+m_{j}+s_{ij})},
\end{equation}
where $T_{ij}$ is the  trips departing  from location $i$ to location $j$, $T_{i}$ is the total  trips departing  from location $i$, $m_{i}$ is the population at location $i$, $m_{j}$ is the population at location $j$, $s_{ij}$ is the total population in the circle of radius $r_{ij}$ centred at location $i$ (excluding the origin $i$ and destination $j$).

\subsection{S{\o}rensen similarity index}
S{\o}rensen similarity index is a statistic tool to identifying the similarity between two samples. It has been widely used for dealing with ecological community data~\cite{sorn}. Ref.~\cite{leno} used a modified version of the index to measure whether  real fluxes are correctly reproduced (on average) by mobility prediction models, defined as
\begin{equation}
\label{eqssi}
SSI \equiv\frac{1}{N^2}\sum^{N}_{i}{\sum^{N}_{j}{\frac{2 \min (T^{'}_{ij},T_{ij})}{T^{'}_{ij}+T_{ij}} }},
\end{equation}
where $T^{'}_{ij}$ is the travels from location $i$ to $j$ predicted by models and $T_{ij}$ is the observed number of trips. Obviously, if each $T^{'}_{ij}$ is equal to $T_{ij}$ the index is 1; if all $T^{'}_{ij}$s are far  from the real values, the index is close to 0.

\section*{Acknowledgments}
We acknowledge the organisers of the D4D Challenge for permitting us to use the Abidjan mobile phone data set. XYY thanks Prof. Ke Xu and Xiao Liang for their enthusiastic sharing of Beijing taxi GPS data. This work was supported by NSFC Grant $61304177$ and $61174150$, Doctoral Fund of Ministry of Education (20110003110027) and partly supported by the opening foundation of Institute of Information Economy, Hangzhou Normal University (PD$12001003002004$).

\newpage

\renewcommand\thesection{S\arabic{section}}
\renewcommand\theequation{S\arabic{equation}}
\renewcommand\thefigure{S\arabic{figure}}
\renewcommand\thetable{S\arabic{table}}
\makeatletter
 \renewcommand\@biblabel[1]{[S#1]}
 \renewcommand\@cite[1]{[S#1]}
\makeatother

\newpage
\setcounter{section}{0}
\setcounter{equation}{0}
\setcounter{figure}{0}

 \begin{LARGE}
{\bf Supplementary Material}
 \end{LARGE}
 
\section{Applying the PWO model to six European cities }

\subsection{Data collection and preprocessing}

\begin{figure}
\begin{center}
\includegraphics[width=9cm]{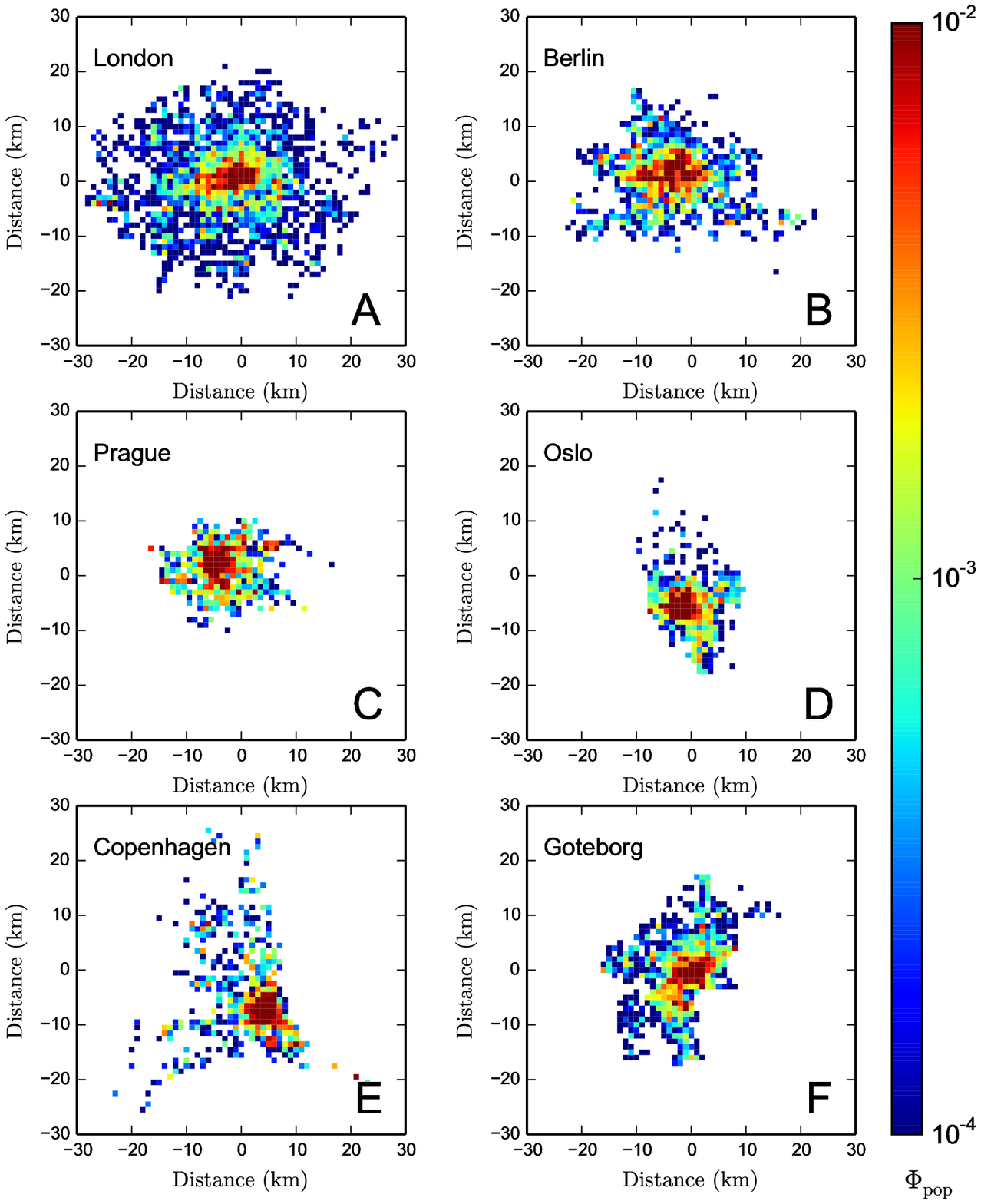}
\caption{\label{fig-map-e}{\bf The zone partition and population density distribution of six European cities}. (A) London. (B) Berlin. (C) Prague. (D) Oslo. (E) Copenhagen. (F) Goteborg. The density function $\Phi _{pop}(i)$ represents the probability of finding a travel started from zone $i$. }
\end{center}
\end{figure}

We use Gowalla check-ins data set~\cite{gw} (downloaded from http://snap.stanford.edu/data/loc-gowalla.html) to test the performance of our population-weighted opportunities (PWO) model in predicting the mobility patterns in European cities. Gowalla is a location-based social networking website in which users share their locations through checking-in. The data set includes in total 6,442,890 check-ins of users over the period of Feb. 2009 - Oct. 2010. For the data set we define a trip of a user's by his/her two consecutive check-ins at different locations. If a trip's origin and destination are both in same city, we classify the trip to inner-city trips. Based on the data, we sort out inner-city trips in six European cities that have a sufficient number of Gowalla users, including London (UK), Berlin (German), Prague (Czech), Oslo (Norway), Copenhagen (Denmark) and Goteborg (Sweden). Table S1 shows the number of trips in each city. According to the methods presented in section {\bf Materials and methods, Data preprocessing} in the main text, we partition each city into equal-area square zones, each of which is of 1 $\rm{km}^2$. Figure~\ref{fig-map-e} shows the zone partition results and the population density distribution of the six cities.

\begin{figure}
\begin{center}
\includegraphics[width=9cm]{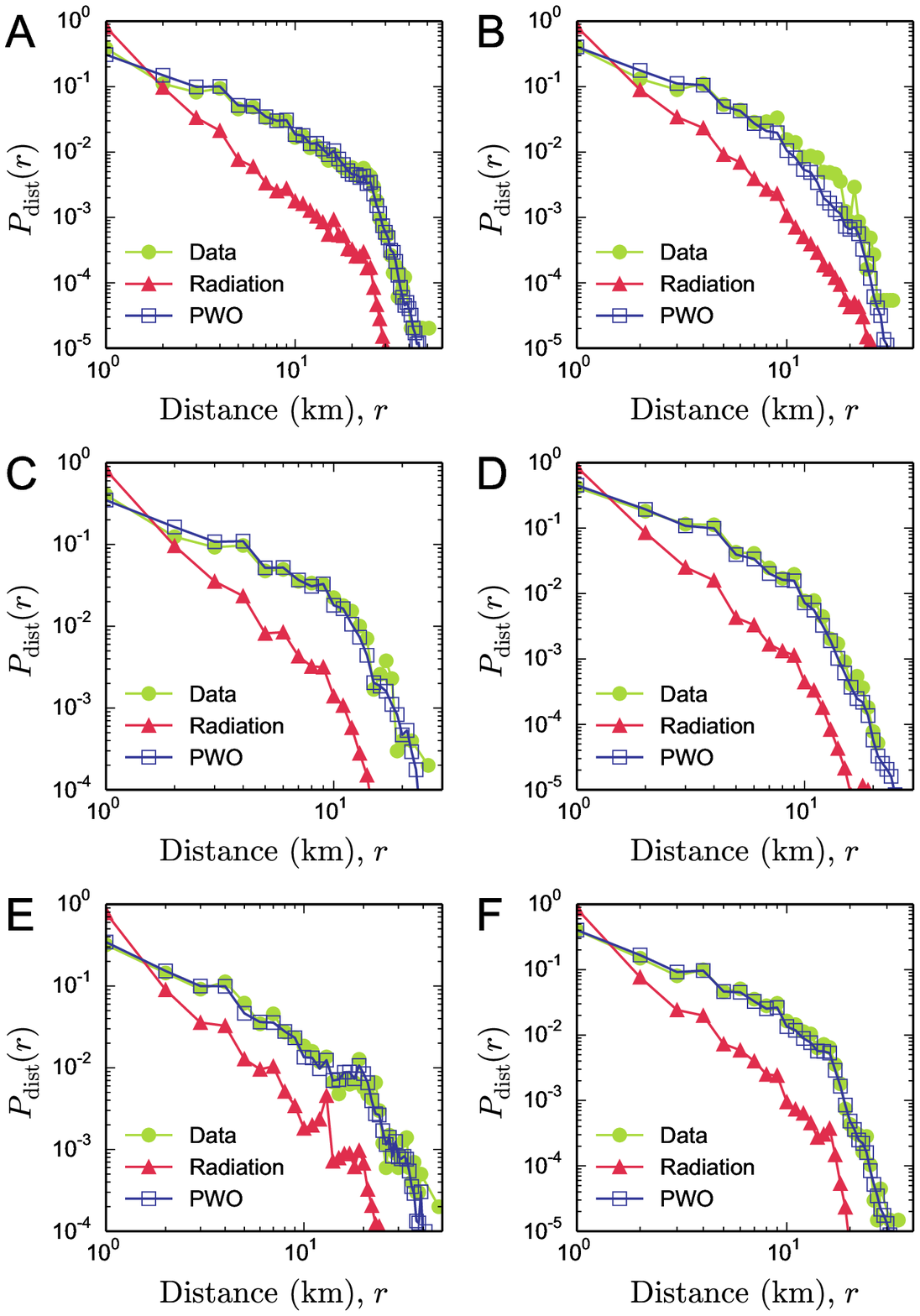}
\caption{\label{fig-tdd-e}{\bf Comparing the travel distance distributions generated by different  models}. (A) London. (B) Berlin. (C) Prague. (D) Oslo. (E) Copenhagen. (F) Goteborg.  $P_{\rm{dist}}(r)$ is the probability of a travel between locations at distance $r$.}
\end{center}
\end{figure}

\begin{figure*}
\begin{center}
\includegraphics[width=18cm]{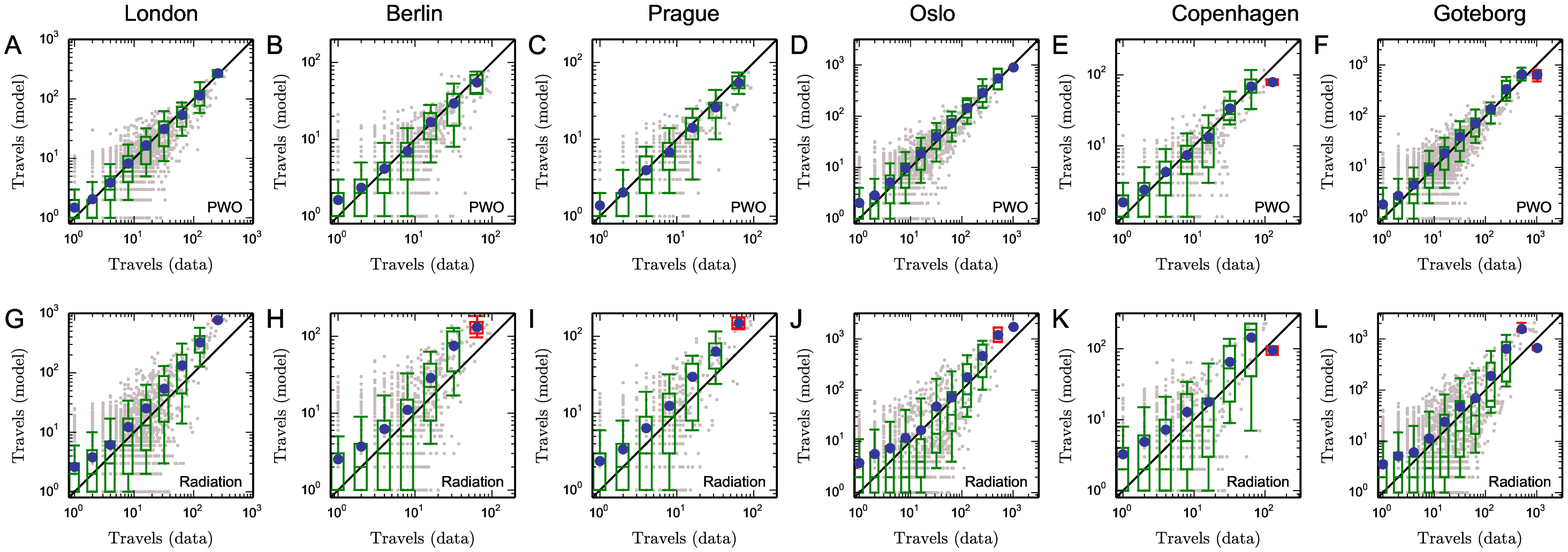}
\caption{\label{fig-box-e}{\bf Comparing the observed fluxes with the predicted fluxes}. (A-F) Predicted fluxes generated by the PWO model. (G-L) Predicted fluxes generated by the radiation model.}
\end{center}
\end{figure*}

\begin{table}[htbp]
    \center
\caption{Data summary of six European cities.}
\begin{tabular*}{0.48\textwidth}{@{\extracolsep{\fill}}lcc}
\hline
\hline{}

City &   Number of Trips & Number of Zones\\
\hline
London  & 49,323 & 986\\
Berlin  & 18,504 & 494\\
Prague & 10,066 & 268\\
Oslo  & 11,464 & 88\\
Copenhagen  & 10,022 & 364\\
Goteborg & 67,438 & 422\\

\hline
\hline
\end{tabular*}
\end{table}

\subsection{Prediction results}

We compare the travel distance distributions and the travel fluxes between all pairs of locations produced by the PWO model and the radiation model. The results are respectively shown in figure~\ref{fig-tdd-e} and figure~\ref{fig-box-e}, in which we can see that the predictions of the PWO model, including the distributions of travel distance and the travel fluxes between all pairs
of locations are in good agreement with real observations, whereas the radiation model's results deviate from the real data. 

\section{Applying the PWO model to four additional U. S. cities }

\subsection{Data collection and preprocessing}

We collected travel survey data from the {\it Metropolitan Travel Survey Archive} website (http://www.surveyarchive.org/), which records more than 40 U. S. cities' travel survey archives. Most surveys contain information of citizens, including their households, vehicles and a diary of their trips on a given day (including each trip's origin and destination location, start and end time, trip mode and purpose). Through checking the data we find that for most data sets the trip-endpoints are labelled by TAZ code or ZIP code. Due to the difficulty in converting the codes to geographic coordinates, we only use the survey data sets of four U. S. cities (New York, Seattle, Detroit and the Twin Cities) that contains the information of trip-endpoint's geographic coordinates (latitude and longitude). 

\begin{figure}
\begin{center}
\includegraphics[width=9cm]{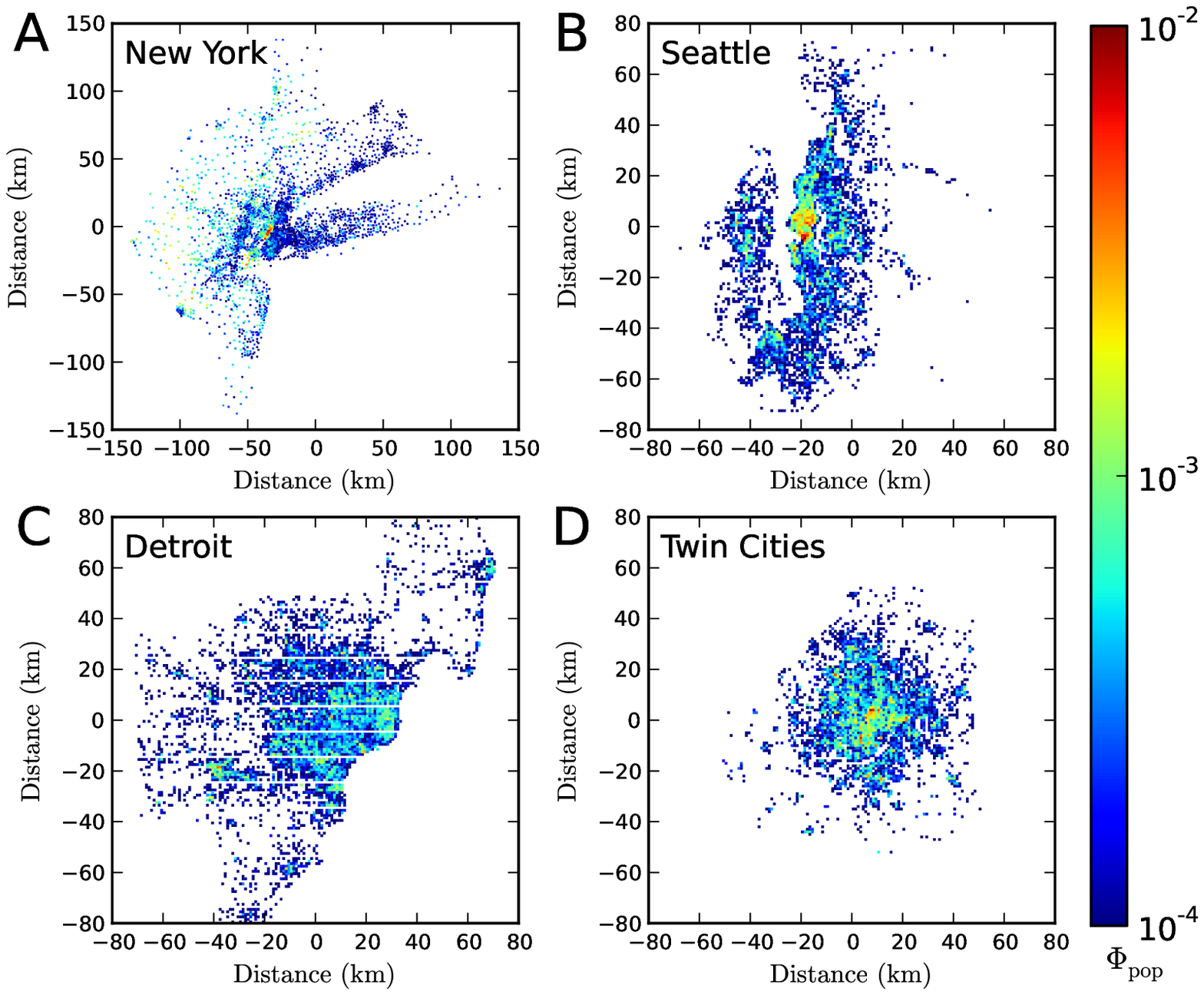}
\caption{\label{fig-map-si}{\bf The zone partition and population density distribution of four U. S. cities}. (A) New York. (B) Seattle. (C) Detroit. (D) Twin Cities. The density function $\Phi _{pop}(i)$ represents the probability of finding a travel started from zone $i$. }
\end{center}
\end{figure}

We find out and record all trips in terms of the coordinates of their origins and destinations from the four data sets. Table S2 shows the number of trips and some other data descriptions of the four cities. Figure~\ref{fig-map-si} shows the zone partition results and the population density distribution of the four cities.

\begin{table}[htbp]
    \center
\caption{Data summary of four U. S. cities.}
\begin{tabular*}{0.48\textwidth}{@{\extracolsep{\fill}}lccccccc}
\hline
\hline{}

City &  Survey Year  &  Households & Trips & Zones \\
\hline
New York& 1998 &10,971&69,282&3,006\\
Seattle& 2006 &4,746&62,277&3,175\\
Detroit& 1994 &7,300&53,583&4,056\\
Twin Cities& 2001 &8,961&35,469&2,684\\

\hline
\hline
\end{tabular*}
\end{table}

\subsection{Prediction results}

\begin{figure}
\begin{center}
\includegraphics[width=9cm]{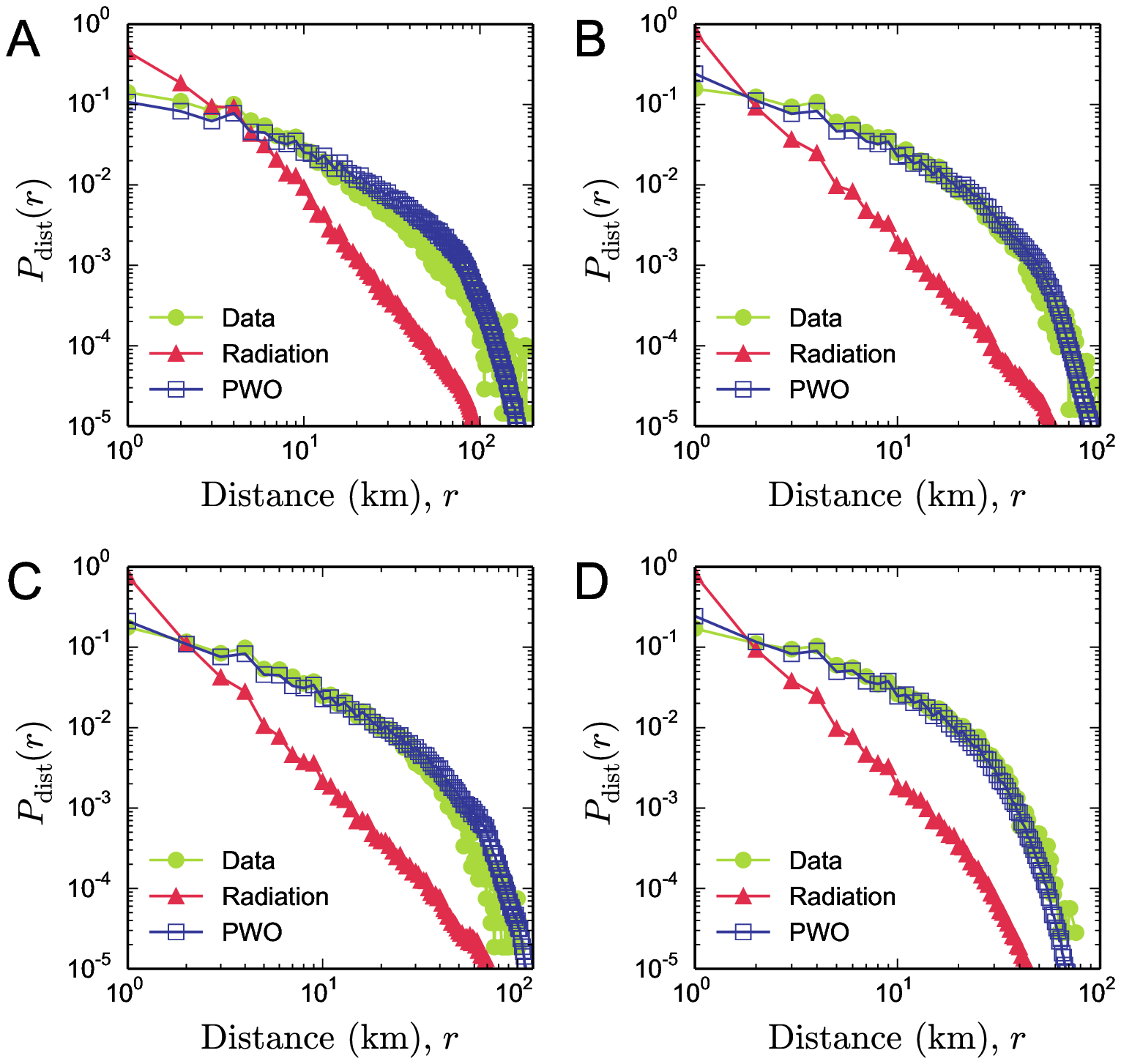}
\caption{\label{fig-tdd-si}{\bf Comparing the travel distance distributions generated by different  models}. (A) New York. (B) Seattle. (C) Detroit. (D) Twin Cities.  $P_{\rm{dist}}(r)$ is the probability of a travel between locations at distance $r$.}
\end{center}
\end{figure}

\begin{figure*}
\begin{center}
\includegraphics[width=18cm]{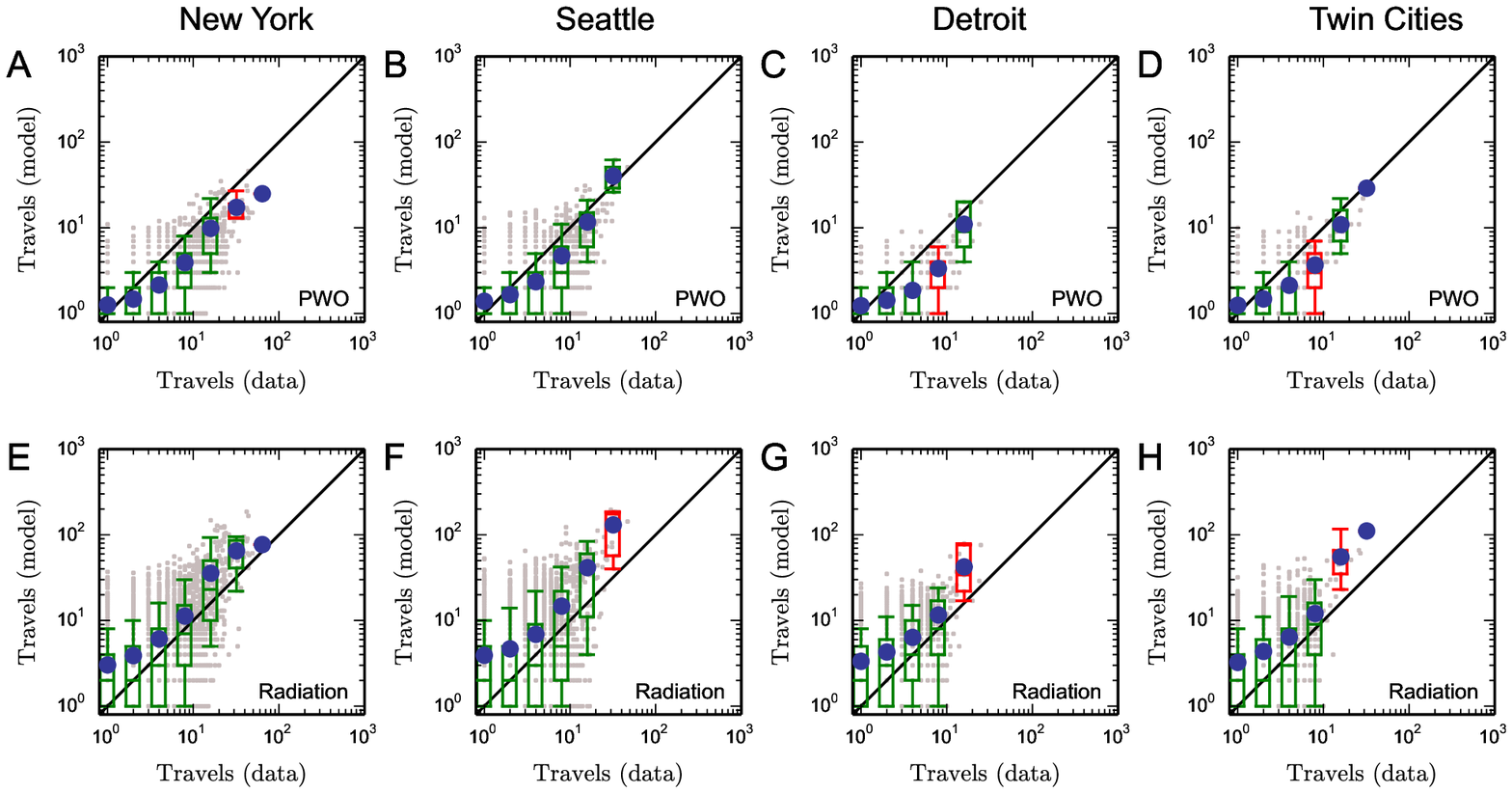}
\caption{\label{fig-box-si}{\bf Comparing the observed fluxes with the predicted fluxes}. (A-D) Predicted fluxes generated by the PWO model. (E-H) Predicted fluxes generated by the radiation model.}
\end{center}
\end{figure*}

Figure~\ref{fig-tdd-si} shows the travel distance distributions produced by the PWO model as well as the radiation model. We can see that the PWO model can precisely reproduce the observed distributions of travel distance, whereas the prediction results of the radiation model deviate from the real data. We also compare the travel fluxes between all pairs of locations produced by both models with the real data. As shown in Figure~\ref{fig-box-si}, we find that the average fluxes predicted by both models deviate from real observations to some extent. The prediction errors result from the low sampling rate in the household travel surveys.
For instance, in New York, there are 3,006 zones and the travel matrix contains more than $9 \times 10^{6}$ elements in principle. However, the surveys only record 69,282 trips. In other word, more than $99\%$ of the real travel matrix's elements are zeros. In contrast, the travel matrices established by the models are always fully filled (although the values of some elements are very small). Thus the fluxes predicted by both models inevitably deviate the insufficient samples of real fluxes. We believe adequate sampling rate of real fluxes will allow a fair comparison between reproduced results and real observations to validate our model.

\section{Comparing the PWO model with parameterised models}
Since the 1940s, many trip distribution models have been proposed for predicting human or freight mobility patterns. The gravity model~\cite{gr} and intervening opportunity model~\cite{io} are two widely used models among them. Since both of them rely on specific parameters estimated in terms of real traffic data to predict mobility model, we name them parameterised models. A recently presented rank-based model~\cite{rank} also belongs to the parameterised models, although it needs very low input information to reproduce some key characteristics of human mobility patterns. In this section, we will compare the prediction performance of the PWO model with those of parameterised models. 

\subsection{The parameterised models}
\noindent {\bf (1) The gravity model}

The gravity model~\cite{gr} stems from Newton's gravity law and has many modified versions so far. The original gravity model has the form
\begin{equation}
\label{eqogr}
T_{ij}=\alpha \frac{ m_{i}m_{j}}{ r_{ij}^{\beta}},
\end{equation}
where $T_{ij}$ is the travels departed from location $i$ to location $j$, $m_{i}$ and $m_{j}$ are the populations of origin and destination and $r_{ij}$ is the distance between $i$ and $j$. Although this model has a very similar form with Newton's gravity law, the model's prediction results may violate the origin constraint $T_{i}=\sum_{j}T_{ij}$ and the destination constraint $T_{j}=\sum_{i}T_{ij}$. To ensure the constraints, one can alternatively use the doubly constrained gravity model~\cite{mt} 
\begin{equation}
\label{eqdgr}
T_{ij}=A_{i} T_{i} B_{j} T_{j} f(r_{ij}),
\end{equation}
where $T_{i}$ is the total travels departed from location $i$, $T_{j}$ is the total travels arrived at location $j$, $f(r_{ij})$ is a function of the distance $r_{ij}$, and $A_{i}=1/\sum_{j}B_{j}T_{j}f(r_{ij})$ as well as $B_{j}=1/\sum_{i}A_{i}T_{i}f(r_{ij})$ are balancing factors that are interdependent to each other. An iterative process enables calculating $A_{i}$ and $B_{j}$, but it demands high computational complexity. To simplify the calculation, one can use the singly constrained versions, either origin or destination constrained, of the gravity model by setting one set of the balancing factors $A_{i}$ or $B_{j}$ equal to one. 

\begin{figure}
\begin{center}
\includegraphics[width=9cm]{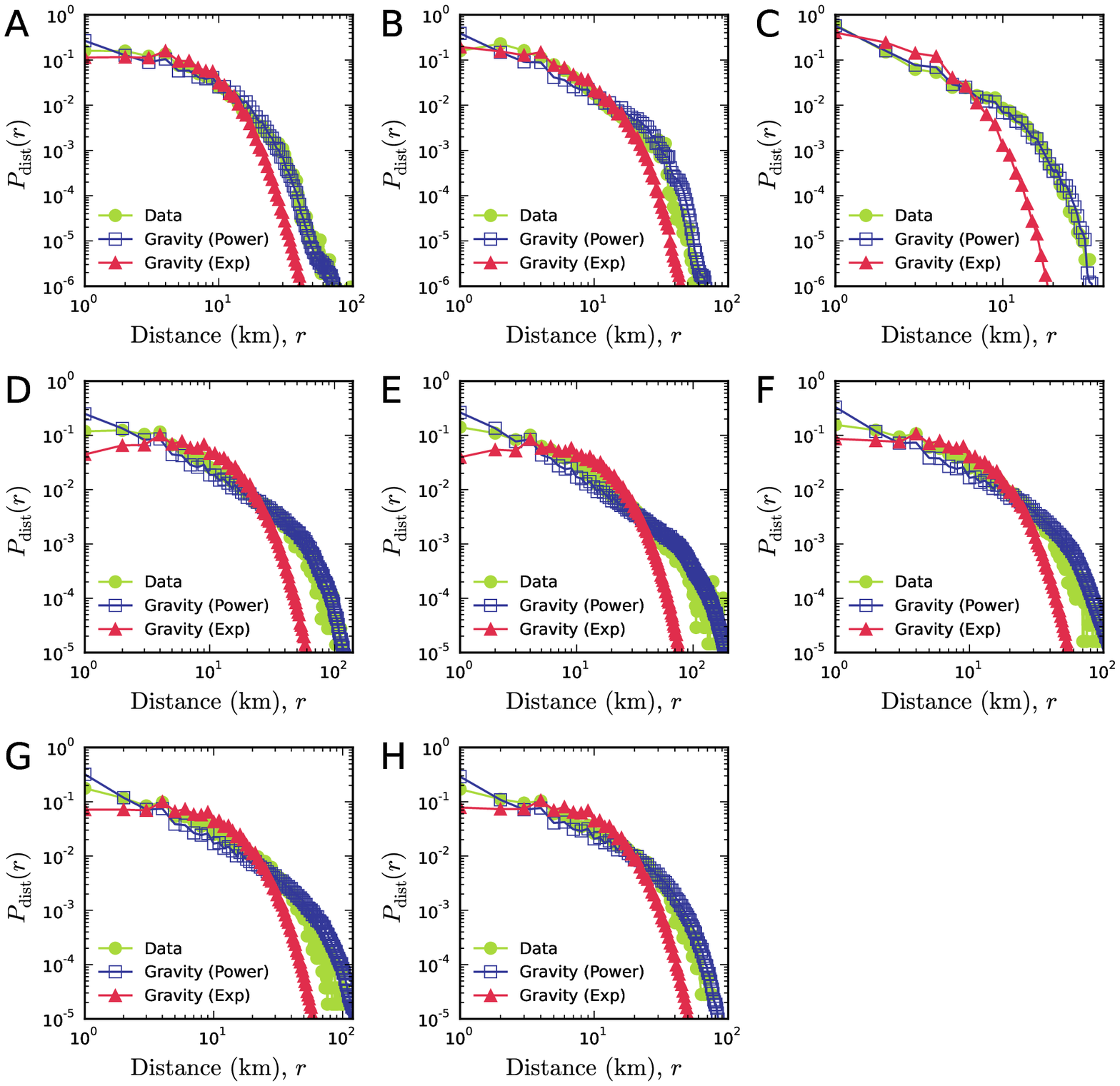}
\caption{\label{fig-tdd-pe}{\bf Comparing the travel distance distributions generated by two types of gravity models}. (A) Beijing. (B) Shenzhen. (C) Abidjan. (D) Chicago. (E) New York. (F) Seattle. (G) Detroit. (H) Twin Cities.}
\end{center}
\end{figure}

\begin{figure*}
\begin{center}
\includegraphics[width=18cm]{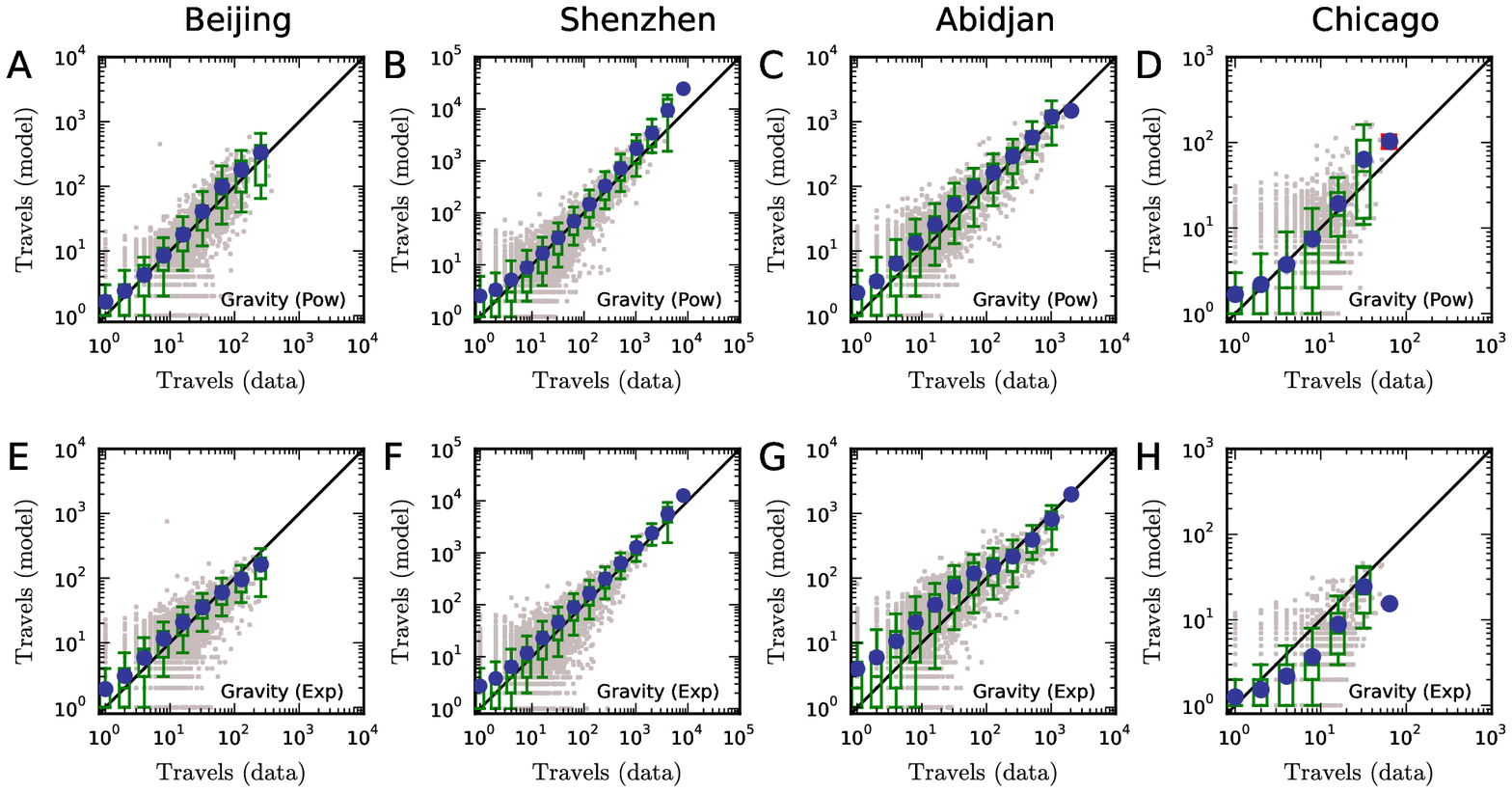}
\caption{\label{fig-box-pe}{\bf Comparing the observed fluxes with the predicted fluxes}. (A-D) Predicted fluxes generated by the gravity model with power-law distance function. (E-H) Predicted fluxes generated by the gravity model with exponential distance function.}
\end{center}
\end{figure*}

Here we employ the origin-constrained gravity model~\cite{mt} to predict mobility patterns in cities, described as
\begin{equation}
\label{eqgr}
T_{ij}=T_{i}\frac{ m_{j}f(r_{ij}) }{ \sum^{N}_{k \neq i} {m_{k}f(r_{ik})} }.
\end{equation}
The distance function $f(r_{ij})$ can be of any forms, such as power or exponential function. Based on numerical test, we find that the gravity model with power function $f(r_{ij})=r_{ij}^{-\beta}$ offers better characterization of the cities' mobility patterns than the exponential function (see figure~\ref{fig-tdd-pe} and figure~\ref{fig-box-pe}). Thus we use the power distance function, the parameter $\beta$ of which is estimated by fitting the real travel data of eight cities.

\noindent {\bf (2) The intervening opportunities (I. O.) model}

The I. O. model~\cite{io} stresses that trip making is not directly related to distance but to the relative accessibility of opportunities for satisfying the objective of the trip. The model's basic assumption is that for every trip departed from a location, there is a constant probability $p$ that determines a traveller being satisfied with a single opportunity. If a location $j$ has $ m_{j}\Theta$ opportunities (we assume the number of opportunities at a location $j$ is proportional to its population $ m_{j}$), the probability of a traveller being attracted by location $j$ is $\alpha m_{j}$, where $\alpha =p \Theta$. 

Considering now the probability $q_{i}^{j}$ of not being satisfied by any of the opportunities offered by the $jth$ destinations away from the origin $i$, we can write 
\begin{equation}
\label{eqiod1}
q_{i}^{j}=q_{i}^{j-1}(1-\alpha m_{j})
\end{equation}
or
\begin{equation}
\label{eqiod2}
\frac{q_{i}^{j}-q_{i}^{j-1}}{q_{i}^{j-1}}=-\alpha m_{j}=-\alpha(S_{i,j}-S_{i,j-1}),
\end{equation}
where $S_{ij}$ is the total population between location $i$ and $j$ (including $i$ and $j$). Assuming that the number of destinations is sufficiently large, we can treat $q$ and $S$ as continuous variables. Then Eq.~(\ref{eqiod2}) can be rewritten as
\begin{equation} 
\label{eqiod3}
\frac{{\rm d}q_{i}}{q_{i}(S)}=-\alpha {\rm d} S.
\end{equation}
After integration we obtain
\begin{equation}
\label{eqiod4}
q_{i}(s) = \frac{e^{-\alpha S}}{1-e^{-\alpha M}},
\end{equation}
where $M$ is the total population in the city. Note that the trip departed from location $i$ to location $j$ is equal to 
\begin{equation}
\label{eqiod5}
T_{ij}=T_{i}[q_{i}(S_{i,j-1})-q_{i}(S_{ij})].
\end{equation}
Combining Eq.~(\ref{eqiod5}) and Eq.~(\ref{eqiod4}), we obtain the I. O. model:
\begin{equation}
\label{eqio}
T_{ij}=T_{i}\frac{e^{-\alpha(S_{ij}-m_{j})}-e^{-\alpha S_{ij}} }{ 1-e^{-\alpha M} }.
\end{equation}

\noindent {\bf (3) The rank-based model}

The rank-based model~\cite{rank} assumes that the probability of an individual travelling from an origin to a destination depends (inversely) only upon the rank-distance between the destination and the origin. The model is described as 
\begin{equation}
\label{eqrk}
T_{ij}=T_{i}\frac{R_{i}(j)^{-\gamma}}{\sum^{N}_{k \neq i} {R_{i}(j)^{-\gamma}}},
\end{equation}
where $R_{i}(j)$ is the rank-distance from location $j$ to $i$ ({\it e.g.}, if $j$ is the closest location to $i$, $R_{i}(j)=1$; if $j$ is the second closest location to $i$, $R_{i}(j)=2$) and $\gamma$ is an adjustable parameter. 

\subsection{Estimating model parameters }

Before applying the parameterised models, it is necessary to estimate their parameters. The goal of the parameter estimation is to maximise the accuracy of reproducing real mobility patterns by the models. Here we use Hyman method~\cite{hym}, a standard method for calibrating gravity model in transportation planning~\cite{mt}, to identify the gravity model's parameter. 

Hyman method aims to find an optimal parameter to minimise the difference between modelled average travel distance and real average travel distance
\begin{equation}
\label{eqerr}
E(\beta)=\left| \bar{r}(\beta) - \bar{r}\right |=\left|  \frac{\sum_{i}{\sum_{j}{T_{ij}(\beta) r_{ij}}}}{\sum_{i}{\sum_{j}{T_{ij}(\beta)}}}-\frac{\sum_{i}{\sum_{j}{T_{ij} r_{ij}}}}{\sum_{i}{\sum_{j}{T_{ij}}}} \right | ,
\end{equation}
where $\bar{r}(\beta)$ is the average distance given by the gravity model with parameter $\beta$, $\bar{r}$ is the real average travel distance, $T_{ij}(\beta)$ is the number of travels from zone $i$ to $j$ generated by the gravity model and $T_{ij}$ is the real number of travels from zone $i$ to $j$. It is not easy to solve the equation $E(\beta)=0$. Hyman suggests that we use the secant method to address this problem, described by the following process: 

\noindent {\bf Step 1}. Give an initial estimate of $\beta_{0} = 1/\bar{r} $.

\noindent {\bf Step 2}. Calculate a trip matrix using the gravity model with the parameter $\beta_{0}$ and obtain a modelled average travel distance $\bar{r}(\beta_{0})$. Estimate a better value of $\beta$ by means of 
\begin{equation}
\beta_{1}=\beta_{0}\bar{r}(\beta_{0})/\bar{r}.
\end{equation}
\noindent {\bf Step 3}. Applying the gravity model with the estimated value of $\beta$ to calculate a new trip matrix and obtain a newly modelled average travel distance $\bar{r}(\beta)$ to compare with $\bar{r}$ . If they are sufficiently close to each other, terminate the iteration and accept the newest value of $\beta$ as the best estimation; otherwise go to step 4. 

\noindent {\bf Step 4}. Improve the estimation of $\beta$ via:
\begin{equation}
    \beta_{i+1}=\frac{(\bar{r}-\bar{r}(\beta_{i-1}))\beta_{i}-(\bar{r}-\bar{r}(\beta_{i}))\beta_{i-1}}{\bar{r}(\beta_{i})-\bar{r}(\beta_{i-1})}.
\end{equation}
\noindent {\bf step 5}. Repeat steps 3 and 4 until $\bar{r}(\beta)$ is sufficiently close to $\bar{r}$.

The estimated parameters of the gravity model, the I.O. model and the rank-based model by Hyman method are listed in Table S3 for different cities. 

\begin{table}[htbp]
    \center
\caption{Estimation of parameter values in three parameterised models}
\begin{tabular*}{0.48\textwidth}{@{\extracolsep{\fill}}lccccccc}
\hline
\hline{}

City &  Gravity   & I. O.  & Rank-based  & Avg. travel dis.\\
 &  $\beta$ & $\alpha$ & $\gamma$ & $\bar{r}$  (km) \\
\hline
Beijing&1.71&5.71$\times 10^{-6}$&1.14&5.50\\
Shenzhen&1.63&3.41$\times 10^{-6}$&1.19&4.77\\
Abidjan&2.43&3.04$\times 10^{-5}$&1.36&2.47\\
Chicago&2.14&2.69$\times 10^{-4}$&1.15&9.49\\
New York&2.28&5.45$\times 10^{-4}$&1.20&11.71\\
Seattle&1.93&2.65$\times 10^{-4}$&1.12&8.30\\
Detroit&2.02&4.59$\times 10^{-4}$&1.13&8.94\\
Twin Cities&1.93&3.56$\times 10^{-4}$&1.04&8.15\\
\hline
\hline
\end{tabular*}
\end{table}

\subsection{Comparison among different models}

\begin{figure}
\begin{center}
\includegraphics[width=9cm]{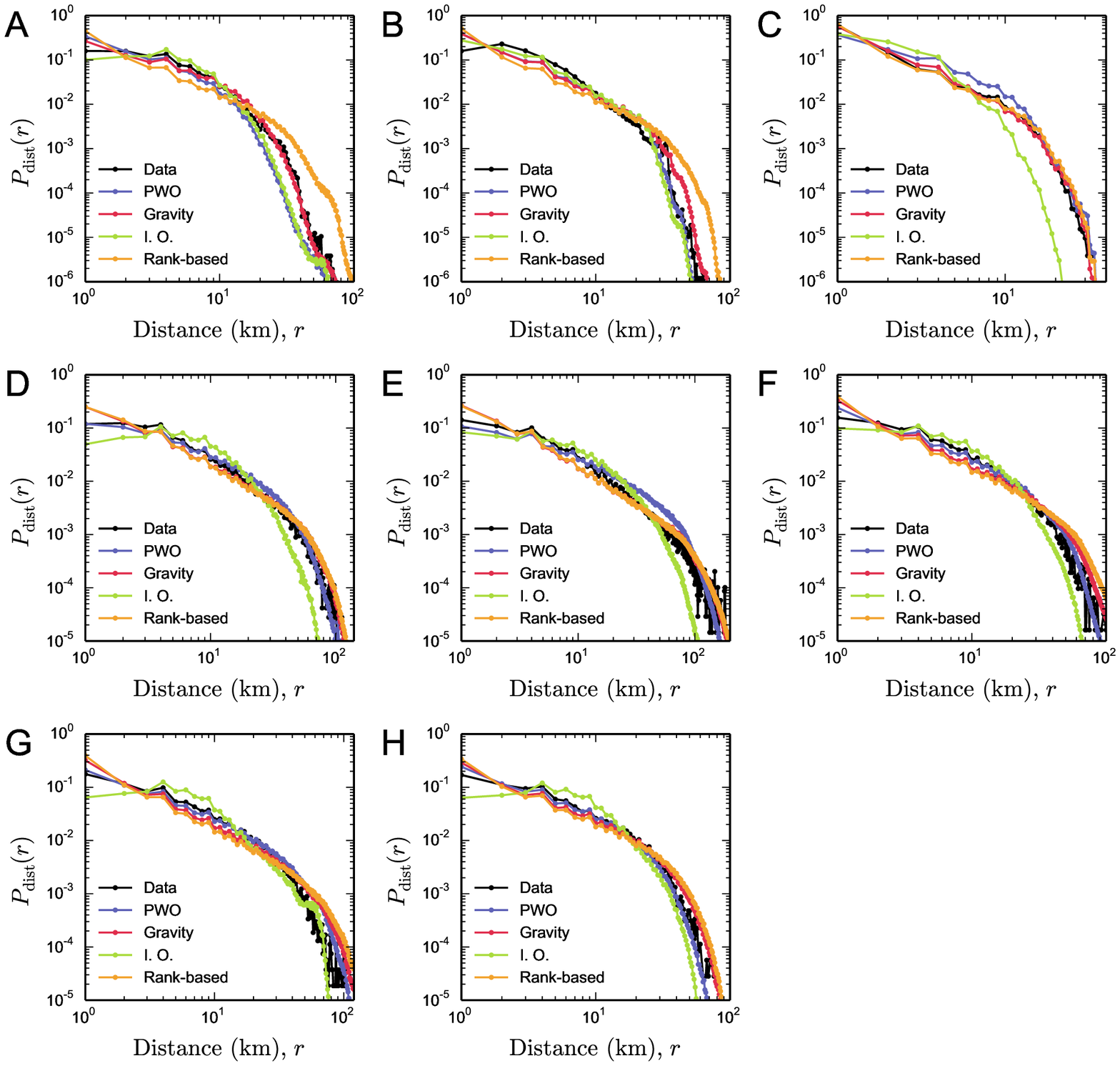}
\caption{\label{fig-tdd-all}{\bf Comparing the travel distance distributions generated by different  models}. (A) Beijing. (B) Shenzhen. (C) Abidjan. (D) Chicago. (E) New York. (F) Seattle. (G) Detroit. (H) Twin Cities.}
\end{center}
\end{figure}

\begin{figure}
\center
\includegraphics[width=9cm]{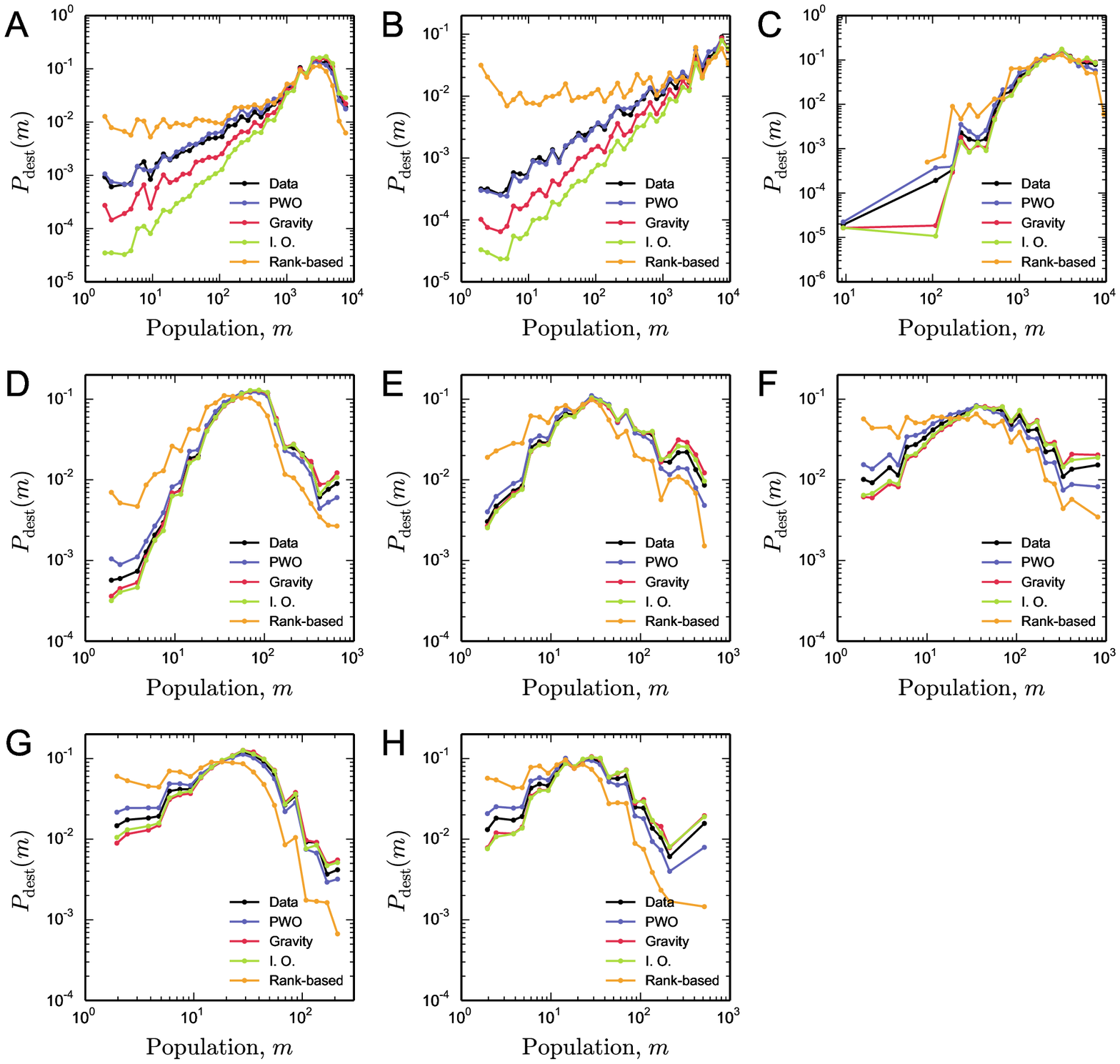}
\caption{\label{fig-des-all}{\bf Comparing the destination travel constraints of different models}. (A) Beijing. (B) Shenzhen. (C) Abidjan. (D) Chicago. (E) New York. (F) Seattle. (G) Detroit. (H) Twin Cities.}
\end{figure}

\begin{figure*}
\centering
\includegraphics[width=18cm]{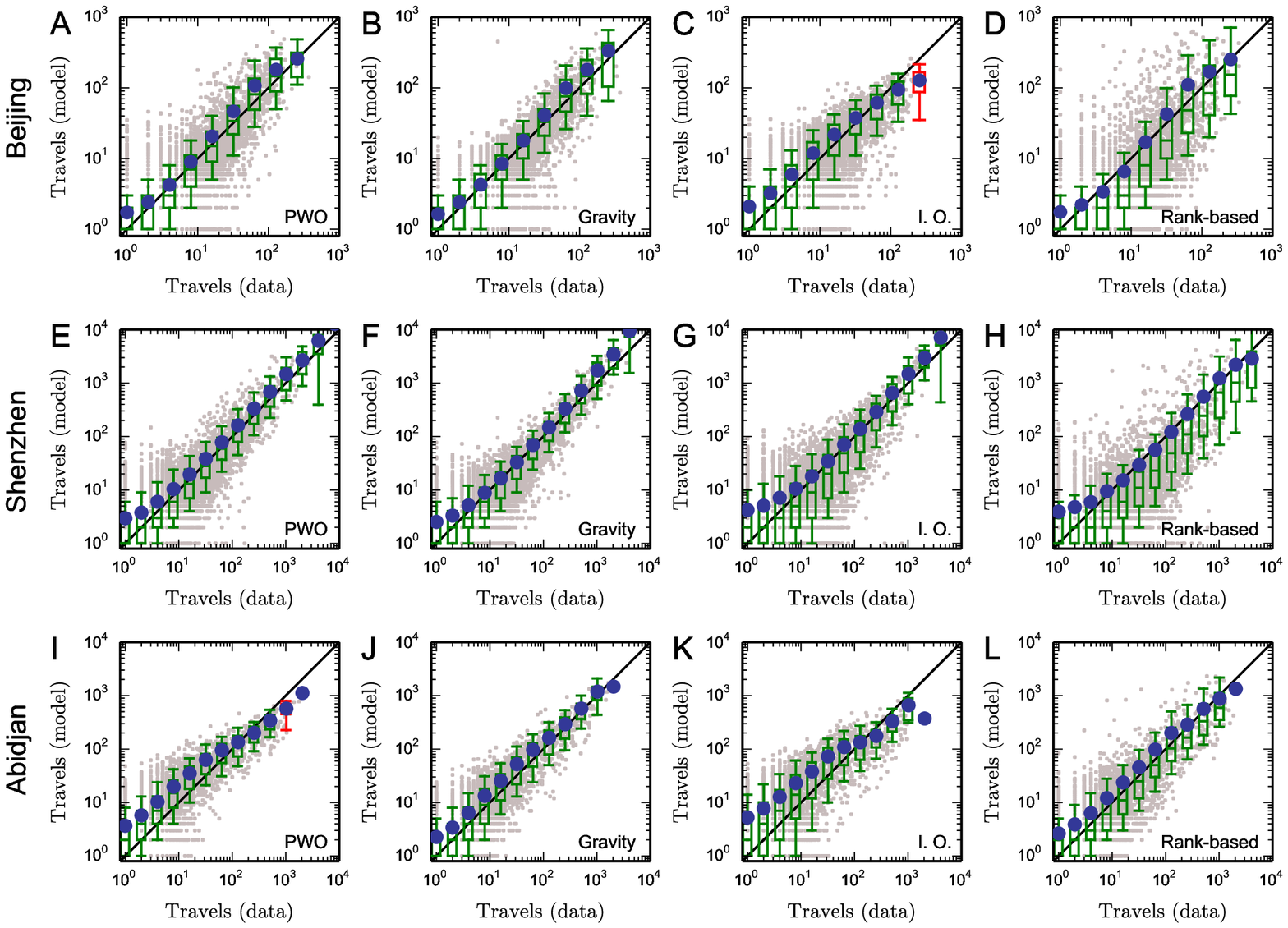}
\caption{\label{fig-box-ch}{\bf Comparing the observed fluxes with the predicted fluxes for three Asian and African cities}. }
\end{figure*}

\begin{figure*}
\center
\includegraphics[width=18cm]{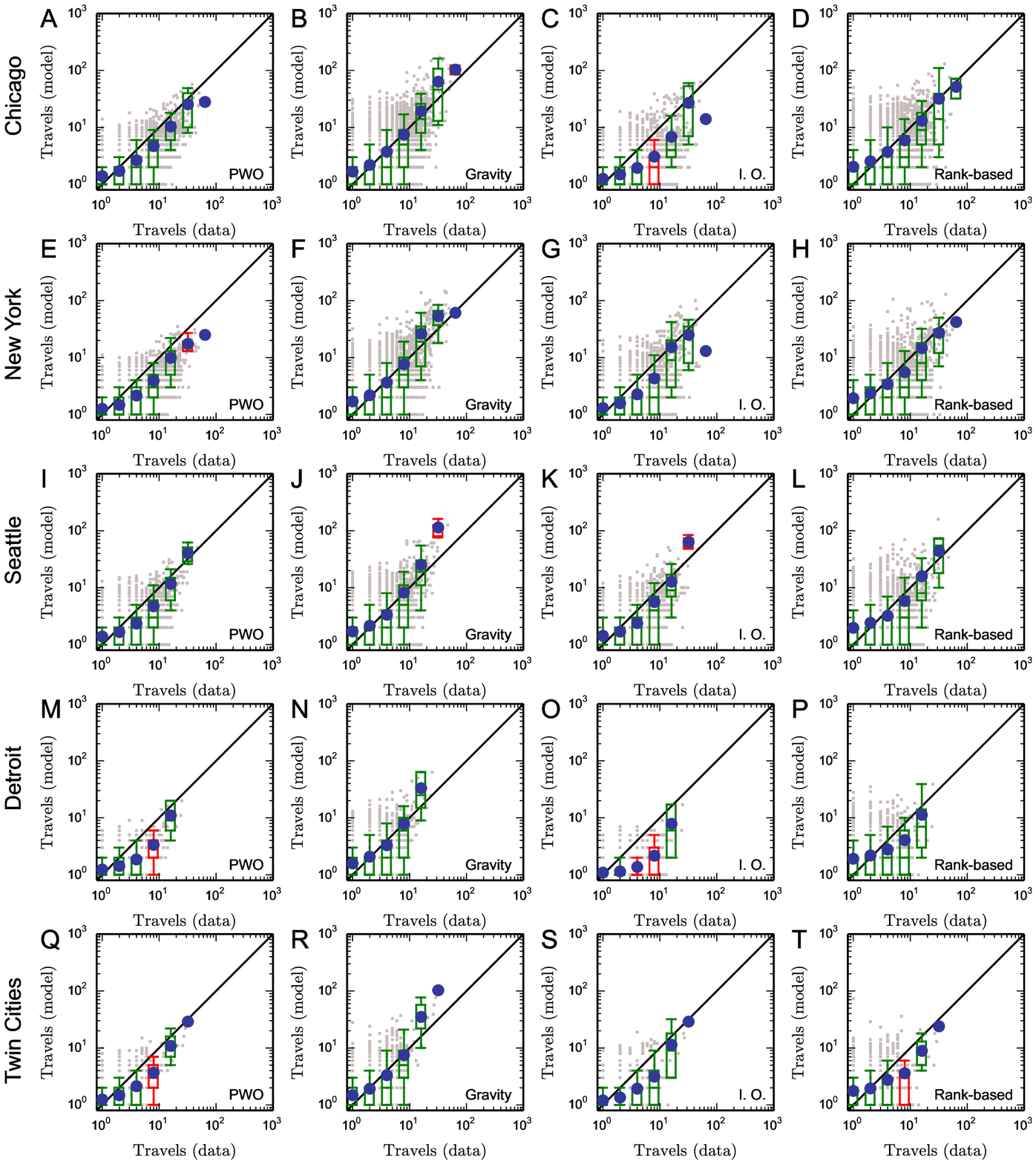}
\caption{\label{fig-box-us}{\bf Comparing the observed fluxes with the predicted fluxes for five U. S. cities}. }
\end{figure*}

\begin{figure}
\center
\includegraphics[width=9cm]{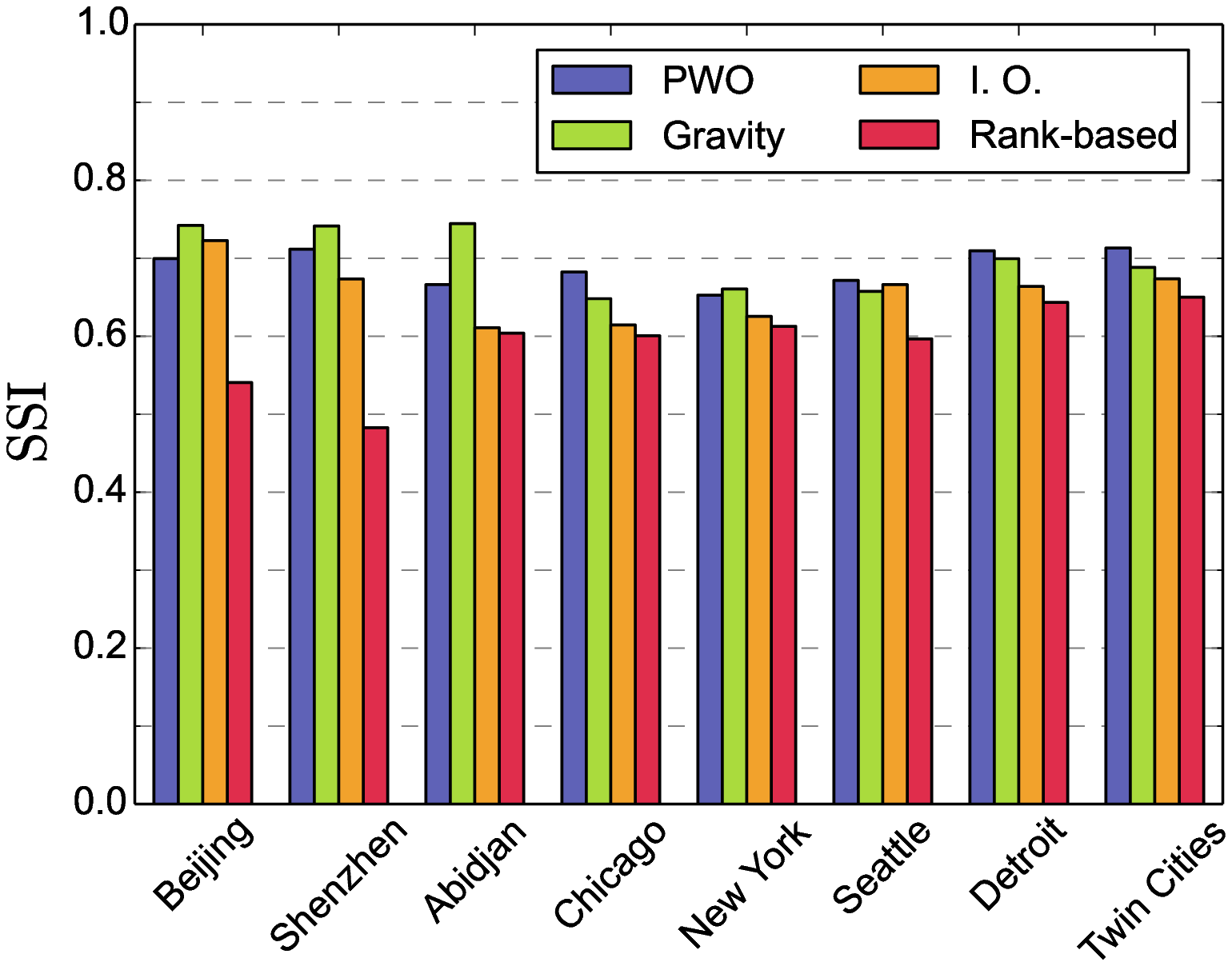}
    \caption{\label{fig-ssi-all}{\bf Comparing the prediction ability of different models based on S{\o}rensen similarity index (SSI)}.}
\end{figure}

\noindent {\bf (1) Travel distance distribution}

Figure~\ref{fig-tdd-all} shows the travel distance distribution predicted by different models. We see that both the gravity model and the rank-based model can reproduce the observed distributions of travel distance in most cases, but the I. O. model's results have significant deviation from the real data. Although in some cases the results of the gravity model (or rank-based model) are better than those of the PWO model, the gravity model needs a distance-decay function with adjustable parameters to match real data, whereas our model relies solely on the population distribution without free parameters. 

\noindent {\bf (2) Destination travel constraints}

Figure~\ref{fig-des-all} shows the destination travel constraints produced by the models. We see that the results from the PWO model are in equal or better agreement with the real data than those of the other models in all cases. It is noteworthy that the rank-based model shows the worst performance in this aspect. We speculate that the fact that the rank-based model do not make use of the population data at destination locations accounts for the big difference from real observations. 

\noindent {\bf (3) Travel fluxes between all pairs of locations}

We compare the travel fluxes between all pairs of locations predicted by models with empirical data. As shown in figure~\ref{fig-box-ch} and figure~\ref{fig-box-us}, we observe that all the predicted average fluxes by the PWO model, the gravity model and the rank-based model are comparable with the real fluxes to some extent. To give a more explicit comparison among different models, we use S{\o}rensen similarity index as an alternative to measure the degree of agreement between reproduced travel matrices and empirical observations. Figure~\ref{fig-ssi-all} shows that on average, the accuracy of the PWO model is higher than that of the I. O. model and the rank-based model. Although in some cases the gravity model can yield better prediction accuracy than the PWO model, the gravity model needs parameter values estimated from previous mobility measurements. In contrast, the PWO model only require the population distribution as input, rendering its application scope broader. 

\section{Relationship among trip distribution models}

To deepen our understanding of the underlying mechanism in the trip distribution models explored in this paper, we discuss the relationship among them. We will first show that, in the particular case of uniform population distribution, the PWO, the radiation model, the I. O. model and the rank-based model can all transform into gravity-like models. Next, we will show that these models can be classified into two categories of modelling frameworks: sequential selection and global selection. 

\subsection{Uniform population distribution }

\begin{figure}
\center
\includegraphics[width=9cm]{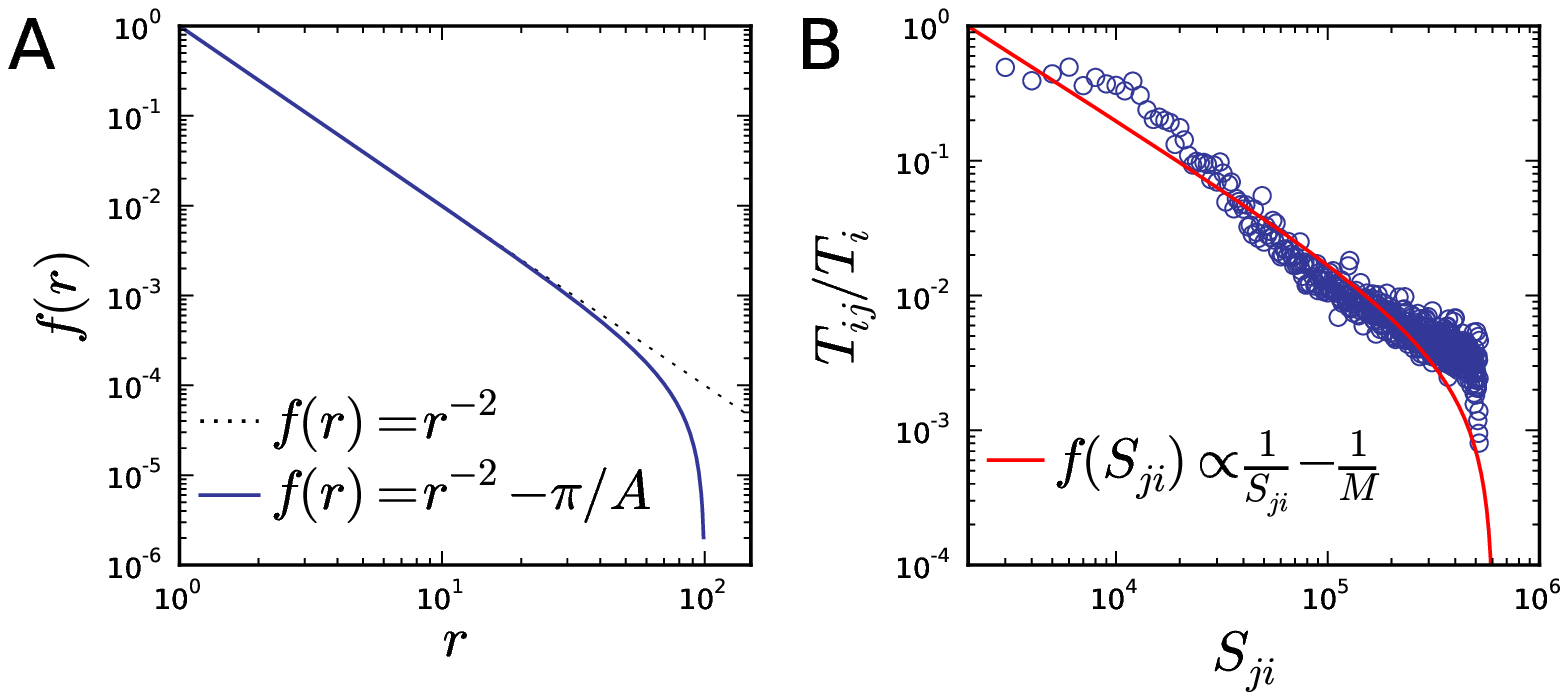}
    \caption{\label{fig-func}{\bf Distance function and population function}. (A) The distance function $f(r_{ij})=r_{ij}^{-2}-\frac{\pi}{A}$. $A=\pi 100^{2}$. (B) The relationship between the population $S_{ji}$ and travel proportion $T_{ij}/T_{i}$. Blue circles are empirical data and red line is the theoretical function ($M=607167$).}
\end{figure}

Consider a uniform population distribution ({\it i.e.} $S_{ji} = \rho \pi r_{ij}^{2}$, where $\rho$ is the population density). We can write the PWO model (Eq.~(1) in the main text) as
\begin{equation}
\label{eqcd1}
T_{ij}=T_{i}\frac{m_{j}(r_{ij}^{-2}-\frac{\pi}{A})}{\sum^{N}_{k \neq i}{m_{k}(r_{ij}^{-2}-\frac{\pi}{A})}},
\end{equation}
where $A$ is the area of the city. Comparing with Eq.~(\ref{eqgr}), we can realise that Eq.~(\ref{eqcd1}) is actually a gravity model with the distance function $f(r_{ij})=r_{ij}^{-2}-\frac{\pi}{A}$. This function is a power law with a cut-off (see figure~\ref{fig-func}(A)). Since the population is not uniformly distributed in real cities, we can not directly use such distance function in the gravity model to predict travel fluxes. Alternatively, we have to estimate its parameters by relying on real traffic data prior to applying the gravity model. However, we may directly choose a population function, such as $f(S_{ji}) = \frac{1}{S_{ji}}-\frac{1}{M}$, to be used in the PWO model in the sense that the heterogeneity of population distribution has been captured by $S_{ji}$. Figure~\ref{fig-func}(B) shows the relationship between the population $S_{ji}$ and the travel proportion $T_{ij}/T_{i}$ in Abidjan, which is in agreement with the population function. 

When the population distribution is of uniform distribution ($s_{ij} = \rho \pi r_{ij}^{2}$), the radiation model~\cite{radi} is reduced to
\begin{equation}
\label{eqrd1}
T_{ij} \propto T_{i}m_{j}r_{ij}^{-4},
\end{equation}
which is actually a gravity model with a power-law distance function with power exponent $\beta=4$. Comparing with the uniform version of the PWO model (having a power exponent $\beta=2$), the selection scope of an individual in the radiation model is relatively more local. The radiation model can characterise the mobility patterns at the country scale (the estimated power exponent of gravity law in the case of U. S. state-wide commuting trips is $3.05$~\cite{radi}, which does not significantly deviate from $4$), but it is not applicable to predicting city mobility patterns. Table S2 shows that the estimated power exponents of the cities are subject to the range $1.63-2.43$, quite close to the exponent in Eq.~(\ref{eqcd1}) but different from that in Eq.~(\ref{eqrd1}). 

The I. O. model can be also transformed into a gravity-like form in the case of uniform population distribution:
\begin{equation}
\label{eqio1}
T_{ij} \propto T_{i} (e^{\alpha m_{j}}-1)e^{-\alpha S_{ij}} = T_{i} (e^{\alpha m_{j}}-1)e^{-\lambda r_{ij}^{2}},
\end{equation}
where $\lambda =\alpha \rho \pi$. The distance function is of high-order exponential form, implying the lack of long-distance travel generated by the model. Thus it is not surprise that the I. O. model usually underestimates long-distance travels, as shown in figure~\ref{fig-tdd-all}. 

The rank-based model uses rank-distance rather than spatial distance to predict the travels between locations. When the population are uniformly distributed in cities, the rank-distance between locations is proportional to the square of the spatial distance, such that the rank-based model can be rewritten as 
\begin{equation}
\label{eqrk1}
T_{ij} \propto T_{i} r_{ij}^{-2 \gamma}.
\end{equation}
The distance function in Eq.~(\ref{eqrk1}) is a power law with the power exponent around 2 (see Table S2). It can thus yield similar results to the travel distance distribution resulting from the gravity model and the PWO model (see figure~\ref{fig-tdd-all}). However, in the rank-based model the information of population in destination is ignored, rendering the destination travel constraints inaccurate, as shown in figure~\ref{fig-des-all}. 

Taken together, insofar as given uniform population distributions, the PWO model, the radiation model, the I. O. model and the rank-based model can presented to be gravity-like models. Although these models have different hypothesis, they share similar underlying mechanism: the probability that an individual selects a travel destination is decreased along with the increment of some prohibitive factors. In the gravity models, the factor is spatial distance; in the rank-based model it is the rank-distance; in the I. O. model, the radiation model or the PWO model, it is the population between origins and destinations. The key difference lies in the fact that the gravity model, the I. O. model and the rank-based model need adjustable parameters to quantify the decrement effect, whereas in the radiation model and the PWO model, the decrement effect is naturally determined by population distribution. 

\subsection{Sequential selection and global selection} 

\begin{figure}
\center
\includegraphics[width=8cm]{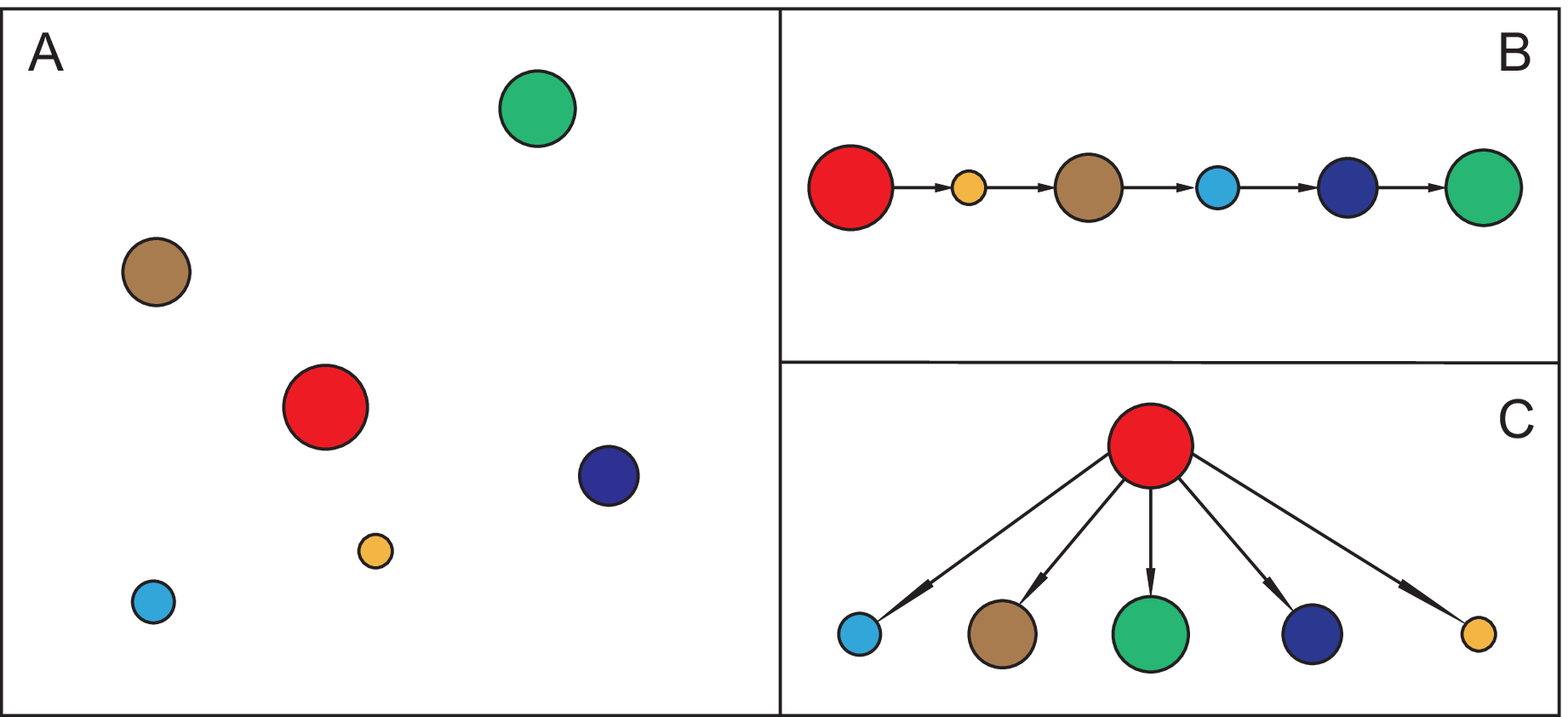}
    \caption{\label{fig-cat}{\bf Schematic description of sequential selection and global selection}. (A) Simplified scenario of destination selection. Central circle (red) is the origin, other circles are selectable destinations. (B) The decision-making process of sequential selection.
    each traveller ranks all possible destinations in ascending order according to the distance to his/her origin. An
    individual first decides whether to travel to the first destination in terms of some probability
    If the individual abandons the first destination, the second one will be considered in terms of the same
probability. Subsequently, all possible destinations will be considered step by step until
the individual decides to travel to the chosen one.
     (C) The decision-making process of global selection. a traveller evaluates the attractions of all possible destinations simultaneously and selects a destination to travel with a probability proportional to the destination's attraction.} 
\end{figure}

According to the decision-making process of travellers for selecting destinations, the frameworks of predicting mobility patterns can be classified into two categories. The first category includes the I. O. model and the radiation model, in which each traveller ranks potential destinations in ascending order according to the distance to his/her origin. An individual first decides whether to travel to the first destination in terms of a probability which is determined by some specific rules (see schematic in figure~\ref{fig-cat}(B)). If the individual abandons the destination, the second one will be considered in terms of the same probability. Analogously, all potential destinations will be considered step by step until the individual eventually decides to travel to a chosen one. We name such step-by-step decision-making process {\it sequential selection}. The modelling framework can be described in a unified form: 
\begin{equation}
\label{eqtt1}
q_{i}^{j}=q_{i}^{j-1}(1-\theta_{j}),
\end{equation}
where $q_{i}^{j}$ is the probability of excluding the 1st to $j$th destinations departed from the origin $i$ and $\theta_{j}$ is the probability of selecting $j$th destination insofar as the 1st to $(j-1)$th destinations are not selected. Thus, the probability of selecting $j$ to travel can be expressed in a joint probability 
\begin{equation}
\label{eqtt10}
p_{ij}=q_{i}^{j-1}\theta_{j}.
\end{equation}

The I. O. model can be derived by assuming that the probability $\theta_{j}$ is proportional to the population of destination $j$ ({\it i.e.} $\theta_{j} = \alpha m_{j}$). After some calculations, we finally have 
\begin{equation}
\label{eqtt11}
p_{ij}=\frac{e^{-\alpha(S_{ij}-m_{j})}-e^{-\alpha S_{ij}} }{ 1-e^{-\alpha M} },
\end{equation}
which is the probability of selecting destination $j$ departed from $i$ in the I. O. model.

Similarly, assuming that the probability $\theta_{j}$ is the ratio of the population of destination $j$ to the total population $S_{ij}$ between locations $i$ and $j$, we have 
\begin{equation}
\label{eqtt2}
q_{i}^{j}=q_{i}^{j-1}(1-m_{j}/S_{ij})=\prod_{j}{\frac{S_{i,j-1}}{S_{ij}}}= \frac{m_{i}}{S_{ij}}
\end{equation}
and
\begin{equation}
\label{eqtt3}
p_{ij}=q_{i}^{j-1}\frac{m_{j}}{S_{ij}}= \frac{m_{i}m_{j}}{S_{i,j-1}S_{ij}},
\end{equation}
which is nothing but the radiation model.

Different forms of probability $\theta_{j}$ can lead to different versions of models that are subject to the framework of sequential selection. 
For instance, assuming the probability $\theta_{j}$ is the ratio of $m_{j}$ to the remaining population $M-S_{ij}+m_{j}$, we can have 
\begin{equation}
\label{eqtt4}
q_{i}^{j}=q_{i}^{j-1}(1-\frac{m_{j}}{M-S_{ij}+m_{j}})= \prod_{j}{\frac{M-S_{ij}}{M-S_{ij}+m_{j}}}= \frac{M-S_{ij}}{M}
\end{equation}
and
\begin{equation}
\label{eqtt5}
p_{ij}=q_{i}^{j-1}\frac{m_{j}}{M-S_{ij}+m_{j}}= \frac{m_{j}}{M},
\end{equation}
which is the uniform selection model~\cite{cont}.

Note that in the sequential selection models, it is possible to find that travellers do not select any destinations to travel unless the system is infinite~\cite{gvr}. In general, the probability is
$p_{ii}=1-\sum_{j}{p_{ij}}$. In other words, 
a traveller stays at the origin with probability $p_{ii}$. For the I. O. model, $p_{ii}=(1-e^{-\alpha m_{i}})/( 1-e^{-\alpha M} )$; for the radiation model and the uniform selection model, $p_{ii}={m_{i}}/{M}$. 

The second category, named {\it global selection}, includes the gravity model, the PWO model and the rank-based model. In global selection model's decision-making process, a traveller evaluates the attractions of all possible destinations simultaneously and selects a destination to travel with a probability proportional to the destination's attraction (see schematic in figure~\ref{fig-cat}(C)). The unified framework can be described as 
\begin{equation}
\label{eqtt6}
p_{ij}=\frac{A_{j}}{\sum_{j}{A_{j}}},
\end{equation}
where $A_{j}$ is the attraction of destination $j$. If we solely use population to capture the destination's attraction, the uniform selection model is obtained; if we use some functions to describe the decay of attraction along with the increase of (real or rank) distance, we can obtain the gravity or the rank-based model (see Eq. (S3) and Eq. (S10)); if the attraction is inversely proportional to the population between destinations and origins, the PWO model is derived. 

Despite the difference between the two modelling frameworks, both sequential and global selection models imply the preference for closer destinations in human travel decision-making: in sequential selection model, a closer destination has a higher priority to be selected; in global selection model, the attraction of a closer destination decays slowly than that of a farther one.
Although both frameworks capture the decision-making process of travellers to some extent, our comparison study (figure S9-S13) demonstrates that at the city scale, global selection models perform better than sequential selection models.

\end{document}